\documentclass[a4paper,12pt]{article}
\pdfoutput=1
\usepackage{graphicx, rotating,amssymb,amsmath,amsfonts,slashed,mathabx,enumerate,soul}
\usepackage[compress]{cite}

\ifx\pdfoutput\undefined
\usepackage[dvips,bookmarks]{hyperref}	
\else
\usepackage{hyperref}	
\fi
\hypersetup{colorlinks,bookmarksopen,bookmarksnumbered,citecolor=magenta,
linkcolor=blus,pdfstartview=FitH,urlcolor=rossos}

\def\hhref#1{\href{http://arxiv.org/abs/#1}{#1}} 

\usepackage{multicol}
\usepackage{color}
\definecolor{rosso}{cmyk}{0,1,1,0.4}
\definecolor{rossos}{cmyk}{0,1,1,0.55}
\definecolor{rossoc}{cmyk}{0,1,1,0.2}
\definecolor{blu}{cmyk}{1,1,0,0.3}
\definecolor{blus}{cmyk}{1,1,0,0.6}
\definecolor{bluc}{cmyk}{1,1,0,0.1}
\definecolor{verde}{cmyk}{0.92,0,0.59,0.25}
\definecolor{verdec}{cmyk}{0.92,0,0.59,0.15}
\definecolor{verdes}{cmyk}{0.92,0,0.59,0.4}

\font\tenrsfs=rsfs10 at 12pt
\font\sevenrsfs=rsfs7
\font\fiversfs=rsfs5
\newfam\rsfsfam
\textfont\rsfsfam=\tenrsfs
\scriptfont\rsfsfam=\sevenrsfs
\scriptscriptfont\rsfsfam=\fiversfs
\def\mathscr#1{{\fam\rsfsfam\relax#1}}

\def\circa#1{\,\raise.3ex\hbox{$#1$\kern-.75em\lower1ex\hbox{$\sim$}}\,}

\newcommand{\beq}{\begin{equation}}
\newcommand{\eeq}{\end{equation}}

\def\circa#1{\,\raise.3ex\hbox{$#1$\kern-.75em\lower1ex\hbox{$\sim$}}\,}
\makeatletter

\def\art{\@ifnextchar[{\eart}{\oart}}
\def\eart[#1]#2#3#4#5#6{{\rm #2}, {#3 #4} {\rm (#6) #5} [{\hhref{#1}}]}
\def\hepart[#1]#2{{\rm #2, \hhref{#1}}}
\newcommand{\oart}[5]{{\rm #1}, {#2 #3} {\rm (#5) #4}}

\newcounter{alphaequation}[equation]
\def\thealphaequation{\theequation\hbox to
0.6em{\hfil\alph{alphaequation}\hfil}}
\def\eqnsystem#1{
\def\@eqnnum{{\rm (\thealphaequation)}}
\def\@@eqncr{\let\@tempa\relax \ifcase\@eqcnt \def\@tempa{& & &} \or
  \def\@tempa{& &}\or \def\@tempa{&}\fi\@tempa
  \if@eqnsw\@eqnnum\refstepcounter{alphaequation}\fi
\global\@eqnswtrue\global\@eqcnt=0\cr}
\refstepcounter{equation} \let\@currentlabel\theequation \def\@tempb{#1}
\ifx\@tempb\empty\else\label{#1}\fi
\refstepcounter{alphaequation}
\let\@currentlabel\thealphaequation
\global\@eqnswtrue\global\@eqcnt=0 \tabskip\@centering\let\\=\@eqncr
$$\halign to \displaywidth\bgroup \@eqnsel\hskip\@centering
$\displaystyle\tabskip\z@{##}$&\global\@eqcnt\@ne
\hskip2\arraycolsep\hfil${##}$\hfil& \global\@eqcnt\tw@\hskip2\arraycolsep
$\displaystyle\tabskip\z@{##}$\hfil
\tabskip\@centering&\llap{##}\tabskip\z@\cr}
\def\endeqnsystem{\@@eqncr\egroup$$\global\@ignoretrue} \makeatother

\def\FERMI{{\sc Fermi}} 
\def\Fermi{{\sc Fermi}} 

\def\HESS{{\sc Hess}}
\def\MAGIC{{\sc Magic}}

\definecolor{darkblue}{rgb}{0,0.2,0.6}
\definecolor{viola}{rgb}{.5,0,.5}
\definecolor{verde}{rgb}{0,.45,0}

\begin{document}
\begin{flushright}
{\footnotesize
{\sc Saclay--t15/130}\\
{\sc Ulb-Phys-th/15-16}
}
\end{flushright}
\color{black}

\begin{center}
{\Huge\bf Gamma ray tests\\[1mm] of Minimal Dark Matter}

\medskip
\bigskip\color{black}\vspace{0.6cm}

{
{\large\bf Marco Cirelli}\ $^a$,
{\large\bf Thomas Hambye}\ $^b$,
{\large\bf Paolo Panci}\ $^c$,\\[3mm]
{\large\bf Filippo Sala}\ $^a$,
{\large\bf Marco Taoso}\ $^a$
}
\\[7mm]
{\it $^a$ \href{http://ipht.cea.fr/en/index.php}{Institut de Physique Th\'eorique}, Universit\'e Paris Saclay, CNRS, CEA,\\ F-91191 Gif-sur-Yvette, France}\\[3mm]
{\it $^b$ \href{http://www.ulb.ac.be/sciences/physth}{Service de Physique The\'orique},
Universit\'e Libre de Bruxelles, \\
Boulevard du Triomphe, CP225, 1050 Brussels, Belgium}\\[3mm]
{\it $^c$ \href{http://www.iap.fr}{Institut d'Astrophysique de Paris}, UMR 7095 CNRS, Universit\'e Pierre et Marie Curie, \\
98 bis Boulevard Arago, Paris 75014, France}\\[3mm]
\end{center}

\bigskip

\centerline{\large\bf Abstract}
\begin{quote}
\color{black}\large
We reconsider the model of Minimal Dark Matter (a fermionic, hypercharge-less quintuplet of the EW interactions) and compute its gamma ray signatures. We compare them with a number of gamma ray probes: the galactic halo diffuse measurements, the galactic center line searches and recent dwarf galaxies observations. We find that the original minimal model, whose mass is fixed at 9.4 TeV by the relic abundance requirement, is constrained by the line searches from the Galactic Center: it is ruled out if the Milky Way possesses a cuspy profile such as NFW but it is still allowed if it has a cored one. Observations of dwarf spheroidal galaxies are also relevant (in particular searches for lines), and ongoing astrophysical progresses on these systems have the potential to eventually rule out the model. We also explore a wider mass range, which applies to the case in which the relic abundance requirement is relaxed. Most of our results can be safely extended to the larger class of multi-TeV WIMP DM annihilating into massive gauge bosons.
\end{quote}

\newpage
\tableofcontents

\bigskip

\section{Introduction}
\label{sec:introduction}

The Large Hadron Collider (LHC) has completed the Standard Model (SM), by unveiling the existence of the higgs boson with a mass of 125.1 GeV. So far, however, the LHC has found no convincing evidence whatsoever for the long-sought-after New Physics at the TeV-scale, expected to be responsible for keeping the mass of the higgs boson so light. Hence, doubts are cast on the very relevance of the {\em naturalness} idea and in particular on the special role of such a scale in modern particle physics. 

On the other hand, thanks to the argument that goes under the name of {\em WIMP miracle} (Weak Interacting Massive Particle), the TeV-scale {\em does} remain appealing in Dark Matter (DM) terms, independently of its possible role or not for the naturalness problem. Indeed, a particle in that mass range and with Weak SM interactions can naturally provide the required DM abundance observed in cosmology via the thermal freeze-out mechanism, which is not a small feat. While this may eventually turn out to be a mirage, it still constitutes a very well motivated driving principle in the quest for the nature of the new particle that has to exist to constitute the observed DM abundance. 

\medskip

Indeed, betting on the possibility of a pure WIMP nature of Dark Matter and of the SM being the ultimate theory all the way up to the GUT or  Planck scale (which is admittedly a bold set of assumptions), the model of Minimal Dark Matter (MDM) was proposed back in 2005~\cite{Cirelli:2005uq,Cirelli:2009uv}. In a nutshell, the construction proposes to add to the SM the minimal amount of new physics (just one extra EW multiplet $\chi$) and search for the minimal assignments of its quantum numbers (spin, isospin and hypercharge) that make it a good DM candidate without ruining the positive features of the SM. No ad hoc extra features are introduced: the stability of the successful candidates is guaranteed by the SM gauge symmetry and by renormalizability. 
By following these principles of consistency to their end and applying the most evident phenomenological constraints (e.g.~direct detection bounds from scattering on nuclei), the theory selects a fermionic $SU(2)_L$ 5-plet with null hypercharge as the {\em only} one which provides, in terms of its electrically neutral component, a viable DM candidate~\footnote{For such a DM candidate no gauge invariant effective operator, made of a DM field and SM fields, hence mediating its decay, can be written with dimension less than 6. A dimension 6 operator induced by the exchange of a GUT scale (or higher scale) heavy particle leads to a large enough DM lifetime, i.e. not ruled out by the constraints which hold on the fluxes of cosmic rays that such a decay scenario would induce. Another originally considered candidate, the scalar 7-plet, is now ruled out by higher dimensional operators (initially overlooked) that mediate its fast decay~\cite{DiLuzio:2015oha,MDM7decay}. In addition, this candidate develops a Landau pole at a low scale~\cite{Hamada:2015bra}.}. The annihilation cross sections of such particles can be fully computed in EW theory, including the Sommerfeld enhancement. Thus, by solving the Boltzmann equation for the relic abundance and requiring such abundance to match the measured cosmological value, the mass of the DM candidate can be univocally determined, turning out to be around 9.4 TeV (we will return in detail on these aspects). 
The components of the multiplet are then split by 1-loop electroweak corrections, which produce small differences.~\footnote{The explicit lagrangian of the model, for reference, reads:
\begin{align}
\mathcal{L}_\chi &= \frac{1}{2}\,\widebar{\chi}(i \slashed{D} - M_\chi) \chi \nonumber \\
& = \frac{1}{2}\,\widebar{\chi_0}(i \slashed{\partial} - M_{\chi_0}) \chi_0 +\widebar{\chi^+}(i \slashed{\partial} - M_{\chi^\pm}) \chi^+ +\widebar{\chi^{++}}(i \slashed{\partial} - M_{\chi^{\pm\pm}}) \chi^{++} \nonumber \\
&+ g  (\widebar{\chi^+}\gamma_\mu \chi^+ + 2\,\widebar{\chi^{++}}\gamma_\mu \chi^{++}) (s_w A_\mu +c_w Z_\mu) \nonumber \\
&+ g (\sqrt{3} \,\widebar{\chi^+}\gamma_\mu \chi_0 + \sqrt{2} \,\widebar{\chi^{++}}\gamma_\mu \chi_+) W_{\mu}^- + {\rm h.c.}
\label{eq:Lagrangian}
\end{align}
where  $\chi$ is the fermion 5-plet Minimal Dark Matter candidate, $g$ is the $SU(2)$ gauge coupling, and $s_w$ and $c_w$ are the sine and cosine of the Weinberg angle. $M_\chi$ is the degenerate mass of the multiplet and $M_{\chi^0}$, $M_{\chi^\pm} = M_{\chi^0} + \Delta M$ and $M_{\chi^{\pm\pm}} = M_{\chi^0} + 4 \Delta M$, with $\Delta M = 166$ MeV, are those of the individual components after splitting.}
This implies that the charged components $\chi^\pm$ and $\chi^{\pm\pm}$ disappear from the thermal bath by decaying, with lifetime $\mathcal{O}({\rm ns})$, into the lightest one $\chi^0$, which constitutes the cosmological DM. 
Due to its minimality, the theory is remarkably predictive: no free parameters are present and therefore the phenomenological signatures can be univocally calculated.

\bigskip

Almost ten years later, and in the context described above, it makes sense to test whether the MDM construction still stands the comparison with data. Broadly speaking: its large predicted mass makes its production cross section at the LHC extremely suppressed~\cite{Ostdiek:2015aga}, and even prospects for future colliders are bleak (see~\cite{Low:2014cba,Cirelli:2014dsa} for related studies); scattering on nuclei for direct detection is much suppressed with respect to initial estimates, as pointed out in a series of recent works (see~\cite{Hisano:2015rsa} and references therein); indirect detection (ID), instead, initially considered in~\cite{Cirelli:2008id,Cirelli:2008jk}, remains promising. 
Among the ID messengers, we focus in this work on gamma rays, and we will comment in the end on other possible channels.

\bigskip

Beyond the motivations specific to the MDM construction, the scale for DM is being pushed towards the TeV and multi-TeV range by null results at the LHC and in direct and indirect detection experiments, as we mentioned above. It seems therefore natural to explore that regime. We will indeed work in a broad range of masses, spanning from 100 GeV to 30 TeV. 
More precisely, we are concerned with three levels of generality in this analysis:
\begin{itemize}
\item[$\diamond$] The specific original MDM 5plet candidate: it has fully determined mass (9.4 TeV) and annihilation cross sections. We will refer to this as `{\em the} MDM 5plet' and point to it with a vertical band in our constraint plots (fig.~\ref{fig:bounds_continuum}, \ref{fig:dwarves} and \ref{fig:lines}). 
\item[$\diamond$] A `generic' 5plet DM candidate: its annihilation cross sections are fixed by EW theory but it can have an arbitrary mass, having relaxed the relic abundance requirement. Namely, for masses smaller (larger) than 9.4 TeV the candidate will be thermally underproduced (overproduced) in cosmology so that alternative mechanisms have to be devised.\footnote{For instance, ref.~\cite{Aoki:2015nza} presents an extension of the MDM model whose net effect is to lower the mass of the DM candidate, even down to less than 1 TeV.} All our plots apply to this case. 
\item[$\diamond$] A generic DM candidate which annihilates into gauge boson channels, such as $W^+W^-$, $ZZ$, $Z\gamma$ and $\gamma\gamma$: most of the bounds that we show can apply to this case.
\end{itemize}

\bigskip

\noindent This paper is organized as follows. 
In Sec.~\ref{sec:sommerfeld} we review the computation of the Sommerfeld enhancement and in Sec.~\ref{sec:relic} that of the relic abundance. 
In Sec.~\ref{sec:constraints} we discuss the constraints from galactic diffuse gamma rays (Sec.~\ref{sec:diffuse}), dwarf galaxies (Sec.~\ref{sec:dwarfs}) and line searches (Sec.~\ref{sec:lines}). In Sec.~\ref{sec:conclusions} we draw a summary and conclude.

\bigskip

Detailed phenomenological analyses of the ID signals of pure $SU(2)_L$ multiplets as DM candidates have been performed in the past, mostly focussed on the case of the triplet, i.e.~the pure Wino (recent examples include~\cite{Cohen:2013ama,Fan:2013faa,Hryczuk:2014hpa,Chun:2015mka,Bhattacherjee:2014dya}).
On the other hand, the case of the 5plet has attracted significant attention in many different contexts recently~\cite{Cai:2011qr,Chang:2012xb,Kumericki:2012bf,Chao:2012mx,Culjak:2015qja,Boucenna:2015haa,Harigaya:2015yaa,DiLuzio:2015oha,Ostdiek:2015aga,Chun:2015mka}, further motivating the analysis that we are embarking on. 
Other recent works with points of contact to ours include~\cite{Cheung:2015mea,Heeck:2015qra}.

\section{Sommerfeld corrections}
\label{sec:sommerfeld}

At non-relativistic velocities, when the mass of the EW multiplet is much larger than the gauge boson masses, certain radiative corrections to the annihilation cross-section can become large, and dominate the lowest order result.
In this regime, the gauge bosons mediate long-range attractive (or repulsive) interactions and the wave functions of the particles involved in the annihilation process become distorted with respect to the plane-waves, enhancing (reducing) the annihilation cross-section.
These effects, the so-called Sommerfeld corrections, can be described as a series of ladder diagrams, where the gauge bosons are exchanged before the annihilation process. 
In practice, the usual pertubation expansion fails, and this class of ladder diagrams has to be resummed.
The Sommerfeld corrections have received a lot of attention during recent years in the context of DM theories, see e.g.~\cite{Iengo:2009ni,Cassel:2009wt,Hisano:2004ds,ArkaniHamed:2008qn,Baumgart:2014saa,Beneke:2014gja,Cirelli:2007xd,Slatyer:2009vg,Schutz:2014nka,Hryczuk:2011vi,Hryczuk:2010zi,Bedaque:2009ri,Bellazzini:2013foa}.
In the following we review how to compute these effects.

\subsection{Sommerfeld factors}

The basic step is the factorization of the short range contribution from the long-range effects, which produce the Sommerfeld corrections. 
The latter ones can be accounted for by computing the wave-function of the system involved in the annihilation, in non-relativistic quantum mechanics.

\smallskip

We classify the initial $\chi\chi$ state with the conserved quantum numbers $Q$, the electric charge, and $S$, the total spin.
We do so because the potential $V$ relevant for the computation depends both on $Q$ and on $S$, so that there are in principle different Sommerfeld effects for different values of $Q$ and $S$.
For definiteness, as well as to make the discussion easier to follow, we often refer to the $Q=0, S=0$ initial state: this is the one relevant for indirect detection signals, because the $\chi_0\chi_0$ initial state indeed has $Q=0$ and $S=0$ (the latter due to its Majorana nature and Fermi statistics). All the other initial states are important for the relic abundance computation, because temperatures in the early universe make all the $\chi$ components on shell, and coannihilations become relevant (as we will discuss in detail in sec.~\ref{sec:relic} ).
\smallskip

An intuitive picture of the computation can be obtained as follows. We are interested in the cross section for the process $a \to {\rm SM \,SM}$, where e.g. $a = \chi_0\chi_0$. We factorize the computation in $a \to i$ and $i \to {\rm SM \,SM}$, where $i$ runs over all the states that can be mixed by the interactions, i.e. those with the same $Q$ and $S$ of the initial state $a$. In the $Q=0, S=0$ case, this amounts to say that $i =1,2,3 = \chi^{++}\chi^{--}, \chi^+\chi^-, \chi_0\chi_0$. The first factor of the computation has to do with the Sommerfeld effect that we describe in this Section, the second factor with the short-distance annihilation process.
\smallskip 

We define $\psi_{ia}$ as the two body non relativistic wave-function in the center of mass frame, where the indexes are defined as above. 
The coupled Schr\"odinger equation governing $\psi_{ia}$ in the center of mass frame is:

\beq 
\label{eq:Schrodinger}
-\frac{\nabla^2}{2 M^r_i}  \psi_{ia} + V_{ij} \psi_{ja} = E \psi_{ia},
\eeq
 
\noindent where $M^r_i$ is the reduced mass of the system and $E$ is defined with respect to the $\chi_0\chi_0$ state: $E = \frac{k_3^2}{2M^r_3}.$  
Here $k_3= 1/2 M_{\chi_0\chi_0} v$ with $v$ the relative velocity of the $\chi_0\chi_0$ particles.
The potential $V_{ij}$ takes into account both the mass splittings and the interactions induced by the gauge bosons. We will give explicit expressions for $V_{ij}$ in Section \ref{sec:VandGamma}.

\smallskip

\noindent The scattering states at large distances have the asymptotic behaviour:

\beq 
\label{eq:asymptotics}
\psi_{ia}(r \rightarrow \infty) = \delta_{ia} e^{i k_i z} + f_{ia}(\theta) \frac{e^{i k_i r}}{r} .
\eeq

\noindent The first term describes the incoming plane-wave wave and $\delta_{ia}$ selects the initial state, for instance for the case of $\chi_0\chi_0$ annihilation $a=3.$ The second part corresponds to the out-going spherical scattering waves. 
The wave-numbers $k_i$ are:

\beq 
\label{eq:ki}
\frac{k_{i}^2}{2 M^r_i} = E - 2\Delta_i,
\eeq

\noindent with $\Delta_i$ the mass splitting, therefore $\Delta_1= 4 \Delta M$,  $\Delta_2 = \Delta M$ and $\Delta_3=0$.

The 3D Schr\"odinger equation can be solved using standard techniques for the scattering problem in quantum mechanics.
We move to a spherical coordinate system and we decompose the wave-function in partial-waves $l$ as:

\beq 
\label{eq:harmonics}
\psi_{ia}= \sum_l \frac{(2l+1)}{k_i} P_l(\cos\theta) \frac{u_{ia}(r)}{r}
\eeq

\noindent where $P_l(\cos\theta)$ are Legendre Polynomials.
The 1D Schr\"odinger equation for the radial function $u(r)$ is:

\beq 
\label{eq:radial}
-\frac{1}{M_{\chi_0}}\frac{d^2}{dr^2} u_{ia}(r) + V_{ij} u_{ja}(r) = E u_{ia}(r)
\eeq

\noindent Imposing the asymptotics of Eq.~(\ref{eq:asymptotics}), the radial function should satisfy:

\beq 
\label{eq:radial_infty}
u_{ia}(r \rightarrow \infty) = \frac{1}{2i} \left( \delta_{ia} e^{ -i k_i r} + S_{ia} e^{i k_i r} \right),
\eeq

\noindent where $S_{ia}$ is closely related to the scattering matrix.
The Schr\"odinger equation admits two sets of solutions, which are either regular around the origin, or singular ${\sim} r^{-l}$ as $r\to 0$.
We can select the regular solutions imposing the asymptotics ($l=0$ for the $s$-wave case under study):
\beq 
\label{eq:bceregular}
u_{ia}(r) \underset{r\to 0}{\sim}  r^{l+1}
\eeq

Summarizing, we should solve Eq.~(\ref{eq:radial}) and impose suitable boundary conditions at the origin and at infinity in order to enforce the asymptotics of Eqs.~(\ref{eq:radial_infty},\ref{eq:bceregular})\footnote{When the $\chi_0\chi_0$ annihilations occour at low velocities, some of the pairs of charged states can be off-shell. For these states $j$ we should set $k_j= i |k_j|$. This  selects the exponentially suppressed mode at infinity.}.
Finally, the Sommerfeld factors $s_{ia}$, which should be convolved with the amplitudes describing the short-range interaction (see later), are simply computed for $s$-wave annihilations taking the value of the wave-function at the origin, see e.g.~\cite{Iengo:2009ni,Cassel:2009wt}.
In terms of the reduced functions $u_{ia},$ they read:

\beq 
\label{eq:sommerfeld}
s_{ia} = \psi_{ia}(0) = \frac{1}{k_i} \frac{d u_{ia}}{dr}(0).
\eeq

The discussion above shows in simple steps how to solve the scattering problem and extract the Sommerfeld factors.
In practice, the differential equation should be solved numerically and some modifications of this method are more appropriate from the point of view of numerical stability.
One possibility is to combine the two sets of solutions of the Schr\"odinger equation, the regular and the singular solutions. 
The details are discussed for instance in~\cite{ArkaniHamed:2008qn,Baumgart:2014saa}. Here we report only the final recipe:

\begin{itemize}

\item[$\diamond$] We solve the coupled Eq.~(\ref{eq:radial}) for each value of $a$ (therefore 3 times for the case of the $Q=0, S=0$ state) with boundary conditions:
\begin{itemize}
\item[i)] $u_{ia}(0)=\delta_{ia}$ \footnote{Here we keep the same notation $u_{ia}(r)$ as before, although now $u$ is not a regular solution of the Schr\"odinger equation.}.

\item[ii)] $u^{\prime}_{ia}(\infty)=i k_i u_{ia}(\infty),$ where the prime stands for the derivative with respect to $r$. This condition originates from imposing the asymptotic behaviour $u_{ia}(r\rightarrow \infty)=R_{ia}e^{i k_i r}$, where the matrix $R_{ia}$ is determined once Eq.~(\ref{eq:radial}) (of course with the boundary conditions i) and ii)) is solved.

\end{itemize}

\item[$\diamond$] The Sommerfeld factors are $s_{ia}= R^{T}_{ia}$, where $T$ denotes the transpose.

\end{itemize}

\noindent We work with the adimensional variable $y=k_3 r.$ The numerical origin is set to $y_{\rm min}=10^{-7}$ and we have checked that smaller values do not affect our results. The numerical infinity is chosen in such a way that a stable solution of the differential equation is achieved.
A different method to solve numerically coupled Schr\"odinger equations have been proposed in~\cite{Ershov:2011zz}, and it has been adopted to compute the Sommerfeld corrections in~\cite{Beneke:2014gja}. We find this method quite efficient for our purposes, and it nicely agrees with the other technique discussed above.

\subsection{Sommerfeld-improved annihilation cross-sections}
\label{sec:VandGamma}

Now we come back to the potential $V$ in Eq.~(\ref{eq:radial}).
It can be derived constructing a non-relativistic field theory for the fermions and integrating out the high energy modes of the spinors and the gauge bosons (see~\cite{Hisano:2004ds}).
In addition to the gauge boson interactions, the potential in Eq.~(\ref{eq:radial}) includes also the mass differences between the different components of the quintuplet.
For the $Q=0$, $S=0$ states, in the basis $\chi^{++}\chi^{--},$ $\chi^{+}\chi^{-}$ and $\chi_0\chi_0,$ one obtains:

\beq 
\label{eq:VS0Q0}
V^{S=0}_{Q=0} = \left(  
	\begin{array}{ccc}
		8\Delta m- 4 A & -2 B & 0\\
		-2 B & 2 \Delta m- A & -3 \sqrt{2} B\\
		0 & -3 \sqrt{2} B & 0
	\end{array}    
 \right)
\eeq

\smallskip

\noindent $A=s_w^2\alpha_{2}/r + \alpha_2 c_w^2 e^{- M_Z r}/r,$ $B=\alpha_2  e^{- M_W r}/r,$ with $\alpha_2=g^2/4\pi.$
One should pay attention to the difference in normalization between the identical and non-identical two particle states, the former ones get in fact an extra $\sqrt{2}$ factor. The factor $\sqrt{2}$ in the off diagonal $_{23}$ entry comes from this mismatch (see e.g.~\cite{Hisano:2004ds}).

Finally, the annihilation cross-section including the Sommerfeld corrections is:

\beq 
\label{eq:sigmatot}
(\sigma v)_{a} = c_a \left( s^{\dagger} \Gamma s \right)_{aa}.
\eeq

\smallskip

\noindent with $s$ the Sommerfeld matrix introduced before, and where the index $a$ is not summed over. The factor $c_a$ is equal to 2 for the $\chi_0 \chi_0$ state ($c_3=2)$ and 1 otherwise.
For the case of $\chi_0 \chi_0$ annihilations we can simply set $a=3.$ 
The matrix $\Gamma$ encodes the short-range annihilation cross-sections. The diagonal entries are simply the $\sigma v$ for the $a$ state into the final state under consideration\footnote{We stress that one should take into account the normalization factors of the quantum-mechanic two-body states, which are different for identical and non-identical particles. The factors $c_a$ are introduced for this reason.}.
Non diagonal entries instead correspond to the quantum interference among  amplitudes involving different initial states. They are imaginary parts of the two-point function of $a \rightarrow b.$ 

\medskip
 
The formalism we have described so far can be applied to compute the annihilation cross-sections of all the components of the quintuplet, in addition to the $\chi_0\chi_0$ initial state that we have mentioned so far. All these processes (annihilations and co-annihilations) are relevant for the calculation of the thermal relic abundance.
The computation can be organised as follow. We label the system with the conserved quantum numbers $Q$ and $S$ and we adopt the basis:

\begin{itemize}

\item[$\triangleright$] $Q=0$, $S=0$: $\chi^{++}\chi^{--},$ $\chi^{+}\chi^{-},$ $\chi_{0}\chi_{0}.$

\item[$\triangleright$] $Q=0$, $S=1$: $\chi^{++}\chi^{--},$ $\chi^{+}\chi^{-}.$

\item[$\triangleright$] $Q=1$, $S=0,1$: $\chi^{++}\chi^{-},$ $\chi^{+}\chi_{0}.$

\item[$\triangleright$] $Q=2$, $S=0$: $\chi^{++}\chi_{0},$ $\chi^{+}\chi^{+}.$

\end{itemize}


The potentials, in agreement with~\cite{Cirelli:2007xd}, are :

\beq 
\label{eq:pot} 
V^{S=1}_{Q=0} = \left(  
	\begin{array}{cc}
		8\Delta m- 4 A & -2 B \\
		-2 B & 2 \Delta m- A 
	\end{array}    
 \right)
\hspace{0.5cm}
V^{S=0,1}_{Q=1} = \left(  
	\begin{array}{cc}
		5\Delta m- 2 A & -\sqrt{6} B \\
		-\sqrt{6} B& 2 \Delta m- 3 B 
	\end{array}    
 \right)
\eeq

\beq 
\label{eq:pot} 
V^{S=0}_{Q=2} = \left(  
	\begin{array}{cc}
		4\Delta m & -2 \sqrt{3}B \\
		-2 \sqrt{3}B & 2 \Delta m+ A 
	\end{array}    
 \right).
\eeq

Using them we can compute the Sommerfeld factors $s_{ia}$ as described in the previous Section. Then, to obtain all the relevant annihilation cross-sections $\sigma v$ from Eq.~(\ref{eq:sigmatot}), one needs the following tree-level $s$-wave annihilation cross sections~\cite{Cirelli:2007xd}:
\beq 
\label{eq:sigmaS0Q0} 
\Gamma^{S=0}_{Q=0} = \frac{\pi \alpha_2^2}{2 M_{\rm DM}^2}
 \left(
    \left(  
	\begin{array}{ccc}
		4 & 10 & 6\sqrt{2} \\
		10 & 25 & 15 \sqrt{2}\\
		 6\sqrt{2} & 15 \sqrt{2} & 18 
	\end{array}    
   \right)
   +
    2 \left(\begin{array}{ccc}
                  16 & 4 & 0 \\
                   4 & 1 & 0 \\
                   0 & 0 & 0
           \end{array}
    \right)
 \right),
\eeq
\beq 
\label{eq:sigmaS0Q12} 
\Gamma^{S=0}_{Q=1} =  \frac{3 \pi \alpha_2^2}{2 M_{\rm DM}^2}
       \left(  
	\begin{array}{cc}
		6 & \sqrt{6} \\
		\sqrt{6} & 1
	\end{array}    
 \right),
  \hspace{.6 cm}
    \Gamma^{S=0}_{Q=2} = \frac{3 \pi \alpha_2^2}{2 M_{\rm DM}^2}
    \left(\begin{array}{ccc}
                  4 & -2 \sqrt{3} \\
                  -2 \sqrt{3} & 3
           \end{array}
    \right),
\eeq
\beq
\label{eq:sigmaS1Q01} 
\Gamma^{S=1}_{Q=0} = \frac{3 \pi \alpha_2^2}{M_{\rm DM}^2}
    \left(\begin{array}{ccc}
                  4 & 2 \\
                  2 & 1
           \end{array}
    \right), 
    \hspace{.6 cm}
    \Gamma^{S=1}_{Q=1} = \frac{3 \pi \alpha_2^2}{M_{\rm DM}^2}
    \left(\begin{array}{ccc}
                  2 & \sqrt{6} \\
                  \sqrt{6} & 3
           \end{array}
    \right),
\eeq
where in (\ref{eq:sigmaS0Q0}) the first matrix refers to the cross section into $WW$, and the second one to the cross sections into all the other gauge bosons. The cross sections into $ZZ$, $\gamma\gamma$ and $Z\gamma$ can be obtained from that second matrix via multiplication by $c_w^4$, $s_w^4$ and $1-s_w^4-c_w^4$ respectively.
\bigskip

\begin{figure}[t]
\begin{center}
\includegraphics[width= 0.48 \textwidth]{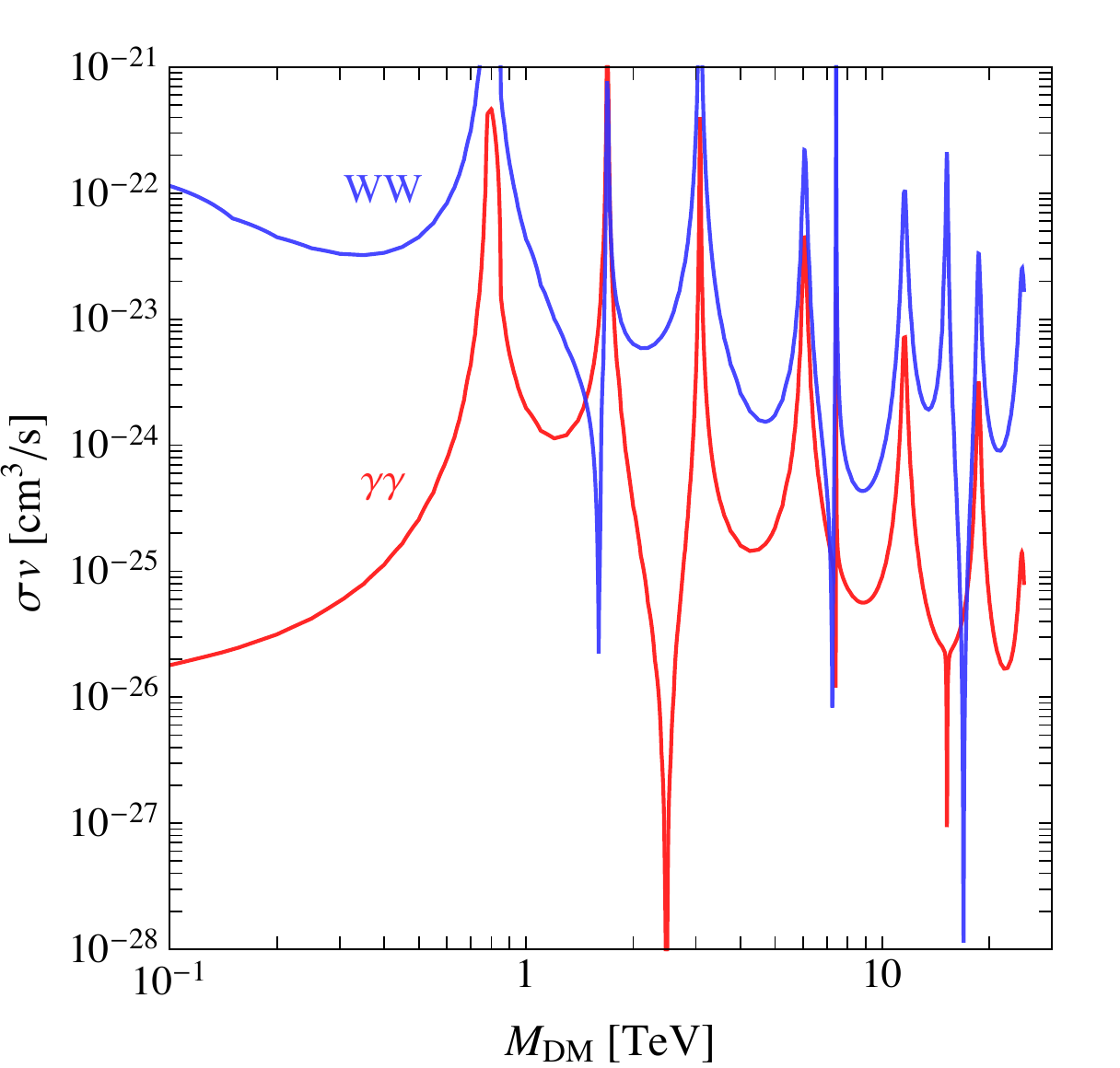} \quad
\includegraphics[width= 0.48 \textwidth]{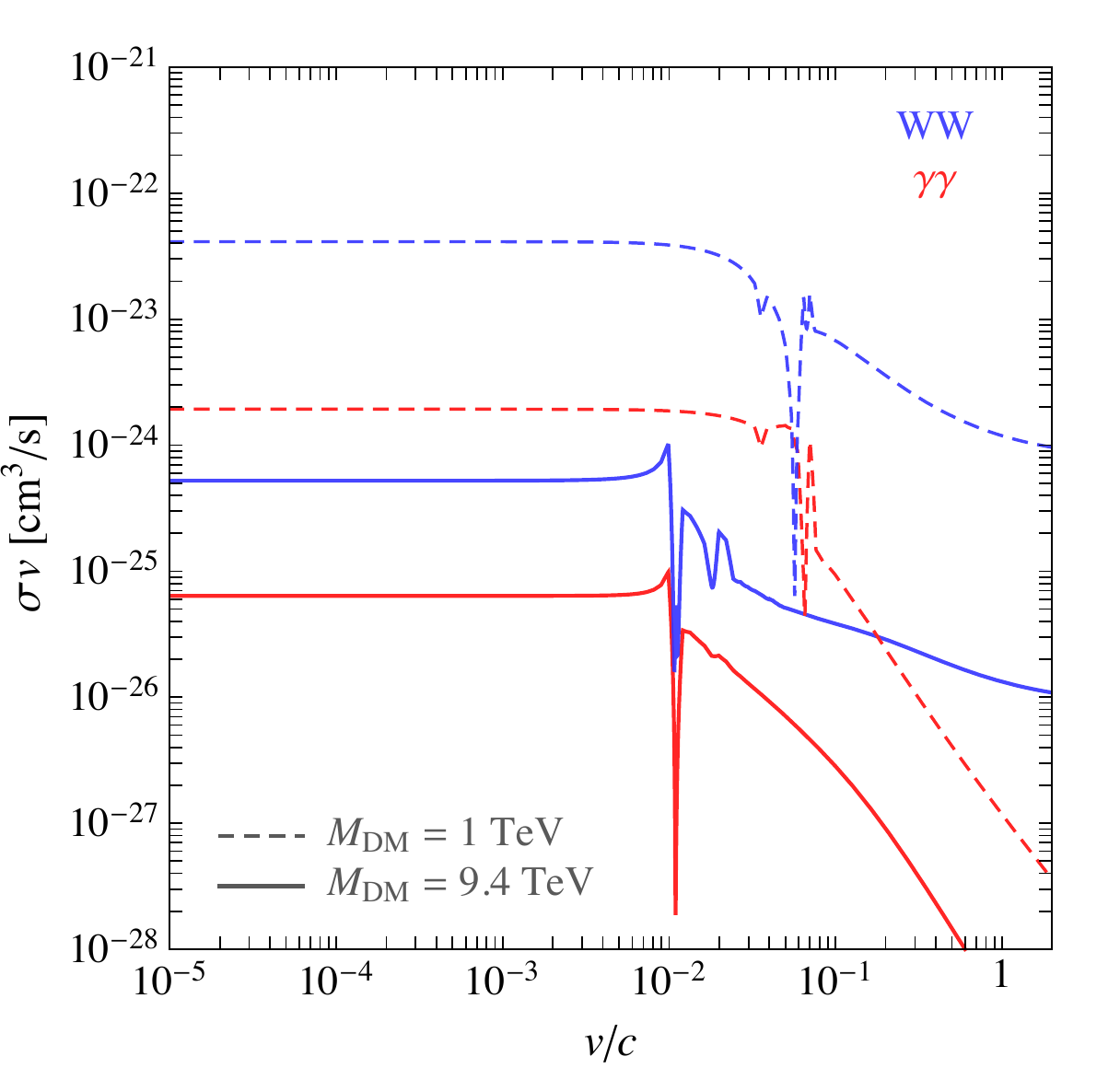}
\caption{\em \small \label{fig:sommerfeldedxsecs} {\bfseries Sommerfeld-enhanced cross sections} for the MDM 5plet, at $v/c = 10^{-3}$ (left, typical of the Milky Way today) and for two fixed values of the MDM mass (right).}
\end{center}
\end{figure}

We conclude this section showing, in Fig.~\ref{fig:sommerfeldedxsecs}, the $\chi_0 \chi_0$ annihilation cross-sections into several final states. For the galactic environment, we fix the typical DM relative velocity to the value $v = 10^{-3} c$.
The series of peaks and dips in the figure are indeed the manifestation of the Sommerfeld effect \footnote{The peaks correspond to bound states while the dips have been interpreted in~\cite{Chun:2012yt} as the `Ramsauer-Townsend effect'.}. 
These results update those obtained in~\cite{Cirelli:2007xd}. We find a good overall agreement, although the precise details of the peaks at large mass (which are important for our purposes) are somewhat modified.

\smallskip
We first notice that the Sommerfeld enhancement can be very big, enhancing the value of the cross sections by several orders of magnitude. Also, while the tree-level short-distance computation would yield $\sigma v = 0$ for the process $\chi_0\chi_0 \to \gamma\gamma$, one observes here a rather large value of this cross section. The presence of such a process is due to the fact that the Sommerfeld effect mixes the $\chi_0\chi_0$ initial state with $\chi^+\chi^-$ and $\chi^{++}\chi^{--}$, which couple to photons.
The fact that the related cross section is rather large is due to the initial state consisting in two doubly charged particles, which results in a factor of 16 enhancement in the cross section, with respect to the analogous process with $\chi^+\chi^-$ instead of $\chi^{++}\chi^{--}$.
We also show the behaviour of the cross sections as a function of the relative velocity, for two given fixed values of the $M_{\rm DM}$. This makes evident that, below a certain threshold which depends on the specific mass, the cross sections stop growing and reach a constant value.
\bigskip


\section{Relic abundance}
\label{sec:relic}
In this section we review the computation of the thermal relic abundance of the lightest component of the 5plet (the Dark Matter particle) as a function of its mass $M_{\rm DM}$. By demanding that it makes all of the measured DM in the Universe ($\Omega_{\rm DM} h^2 = 0.1188\, \pm \, 0.0010$, as determined by the latest {\sc Planck} results~\cite{Ade:2015xua}) we can univocally determine its mass. 

\medskip

In principle, the evolution of the number density $n_\alpha$ of each $\alpha$-th component of the 5plet $\chi_\alpha \equiv (\chi_0,\chi^+,\chi^-,\chi^{++},\chi^{--}$), with  internal degrees of freedom $g_\alpha \equiv 2$ and mass $M_\alpha \equiv M_{\rm DM}+ \Delta M_\alpha$,
has to be computed by solving a system of five coupled Boltzmann equations. 
The densities are affected, in addition to the process of $\chi_0 \chi_0$ annihilation into any SM state, by all the possible $\chi_\alpha \chi_\beta$ co-annihilations. Indeed, since the states are almost degenerate in mass, the $\chi_\alpha\chi_\beta$ co-annihilation cross sections play a major role in setting today's relic abundance. 
However, in practice, it turns out that it is sufficient to follow the evolution of the {\em total} number density  $n=\sum_\alpha n_\alpha$, i.e. it is sufficient to just solve one simple Boltzmann equation in terms of the total thermally averaged annihilation cross section which includes all the possible $\chi_\alpha\chi_\beta$ co-annihilation channels.
This is because $n_\alpha$ is much smaller than the SM thermal bath number density $n_{\rm SM}$~\footnote{The almost degenerate states of the 5plet have  masses $M_\alpha$ much larger than those of the SM particles and therefore $n_\alpha$ is suppressed by a Boltzmann factor, while $n_{\rm SM}$ is not.} and, at the same time, the scattering cross sections $\sigma_{\alpha \, {\rm SM} \, \rightarrow \, \beta \, \rm SM}$ are of the same order of the annihilation cross sections $\sigma_{\alpha  \beta \, \rightarrow \, {\rm SM \, SM}}$. Hence the scattering rates off SM particles are much faster than the $\chi_\alpha\chi_\beta$ annihilation rates and the $\chi_\alpha$ particles are kept in equilibrium with the thermal bath. As a consequence, the only relevant processes are those that modify the total density $n$.
This greatly simplifies the problem. 

As it is customary, we actually define the total comoving  density $Y=n/s$, where $s$ is the total entropy density of the Universe, and we follow the evolution of $Y$ in terms of the adimensional parameter $x=M_{\rm DM}/T$, with $T$  the temperature of the thermal bath. In terms of $x$, the entropy density reads $s(x)\simeq 2\pi^2/45 \, g_{* s}(x) M_{\rm DM}^3 \, x^{-3} $ and the Hubble rate in the radiation domination era reads $H(x)=\sqrt{4\pi^3 g_*(x)/45} \, M_{\rm DM}^2/M_{\rm pl} \, x^{-2}$.  Here $g_{* s}(x)$ and $g_{*}(x)$ are the effective relativistic degrees of freedom as given in Ref.~\cite{Gondolo:1990dk}.

We compute the cross sections as described in Sec.~\ref{sec:sommerfeld}, i.e. combining the Sommerfeld effect with the $s$-wave short distance contributions of Eqs.~(\ref{eq:sigmaS0Q0},\ref{eq:sigmaS0Q12},\ref{eq:sigmaS1Q01}). This procedure has an uncertainty coming from the neglection of the $p$-wave contributions to Eqs.~(\ref{eq:sigmaS0Q0},~\ref{eq:sigmaS0Q12},~\ref{eq:sigmaS1Q01})~\cite{Cirelli:2009uv}, as well as from having neglected the relativistic corrections to the Schr\"odinger equation (\ref{eq:Schrodinger}). We account for these effects by assigning a rough $v^2/c^2$ uncertainty to the thermal relic mass, which is of the order of 5\% at freeze-out.
We have accounted for the dominant thermal corrections (as in~\cite{Cirelli:2007xd}), which however have a very small impact.

\medskip

We can now write the Boltzmann equation for $Y$ as a function of $x$. Following Refs.~\cite{Gondolo:1990dk, Edsjo:1997bg}, it explicitly writes 

\beq\label{Eq:BE}
\frac{{\rm d} Y(x)}{{\rm d} x}=-\frac{s(x) \langle \sigma_{\rm eff} v\rangle }{x H(x)}\left(1-\frac{x}{3 g_{* s}(x) } \frac{{\rm d} g_{*s}(x)}{{\rm d} x}  \right)\left(Y^2(x)-Y_{\rm eq}^2(x) \right) \ ,
\eeq
where $Y_{\rm eq}= 45 \, g_{\rm tot}(x)/(4 \pi^4  g_{*s}(x)) \, x^2 K_2(x) $ is the equilibrium comoving density, $K_2$ is the BesselK function of the second order and   $g_{\rm tot}(x)=\sum_\alpha g_\alpha (1+\xi_\alpha)^{3/2}\,e^{-x\, \xi_\alpha }$, with $\xi_\alpha = \Delta M_\alpha /M_{\rm DM}$, are the total effective degrees of freedom of the 5plet at a given temperature corresponding to $x$. Notice that, in the limit of degenerate states ($\xi_\alpha \equiv 0 $ for all $i$),  $g_{\rm tot}(x)=\sum_\alpha g_\alpha = 10$. The initial condition of $Y$ in Eq.~\eqref{Eq:BE} reads as usual $Y(x_0)=Y_{\rm eq}(x_0) $, where we  fix $x_0=4$ in order to follow the whole subsequent evolution. 

In Eq.~\eqref{Eq:BE}, $\langle \sigma_{\rm eff} v\rangle $ is the total thermally averaged effective annihilation cross section in 
the cosmic comoving frame, which is the most important quantity to determine the today's relic abundance of the MDM 5plet. Under the reasonable assumption of thermal equilibrium of the $\chi_\alpha$ state with the SM thermal bath, as discussed above, it writes~\cite{Servant:2002aq} 

\begin{eqnarray}
\langle \sigma_{\rm eff} v\rangle&=&\sum_{\alpha,\beta} \, \frac{g_\alpha g_\beta}{g_{\rm tot}^2(x)}\langle \sigma_{\alpha \beta} v\rangle (1+\xi_\alpha)^{3/2} (1+\xi_\beta)^{3/2} e^{-x\left(\xi_\alpha+\xi_\beta \right)} \ ,  \label{Eq:sigmaeff} \\ 
\langle \sigma_{\alpha\beta} v\rangle& = & \frac{x}{8M_{\rm DM}^5 K_2^2(x)} \int_{4 M_{\rm DM}^2}^\infty \hspace{-.25cm}{\rm d} s \,  \sigma_{\alpha\beta}(s) \, (s-4M_{\rm DM}^2)\sqrt s \, K_1\left(\frac{\sqrt s \, x}{M_{\rm DM}}\right) \ , \label{Eq:sigmavij}
\end{eqnarray}
where $K_1$ is the BesselK function of the first order and the $\sigma_{\alpha\beta}$ are all the possible annihilation cross sections into any SM state  (see Sec.~\ref{sec:sommerfeld}).\footnote{The single particle indices $\alpha$ and $\beta$ running over the components of the multiplet were collectively denoted with the 2-particle index $a$ in the previous Section and in particular in Eq.~(\ref{eq:sigmatot}).} Since, for velocity dependent cross sections (e.g. the ones which include the Sommerfeld corrections) and large $x$, the integral in Eq.~\eqref{Eq:sigmavij} is numerically hard to solve, for $x>100$ we introduce the variable $\epsilon=1/x$ and  we expand the function $K_1(\sqrt s /(\epsilon\, M_{\rm DM}))/K_2^2(1/\epsilon)$ for small $\epsilon$. We have checked that this approximation is very accurate by comparing the exact integral in Eq.~(\ref{Eq:sigmavij}) with the approximated one, for velocity independent cross sections. Considering all the possible CP combinations of the initial states, Eq.~\eqref{Eq:sigmaeff} in the limit of degenerate masses explicitly writes 
\beq
\begin{split}
\langle \sigma_{\rm eff} v\rangle  = \frac1{25} & \left[ \langle \sigma v \rangle_{0\, 0}^{S=0}  +   
2 \left(\langle \sigma v \rangle_{+\,-}^{S=0} + \langle \sigma v \rangle_{+\,-}^{S=1}\right) + 2 \left(\langle \sigma v \rangle_{++\,--}^{S=0} + \langle \sigma v \rangle_{++\,--}^{S=1}\right) + \right. \\
& \left. 4 \left(\langle \sigma v \rangle_{+\,0}^{S=0} + \langle \sigma v \rangle_{+\,0}^{S=1}\right) + 4 \left(\langle \sigma v \rangle_{++\,-}^{S=0}+ \langle \sigma v \rangle_{++\,-}^{S=1}\right) + 2 \,  \langle \sigma v \rangle_{+\,+}^{S=0} + 4 \,  \langle \sigma v \rangle_{++\,0}^{S=0} \right] \ .
\end{split}
\eeq 

\medskip

\begin{figure}[t]
\begin{center}
\includegraphics[width= 0.475 \textwidth]{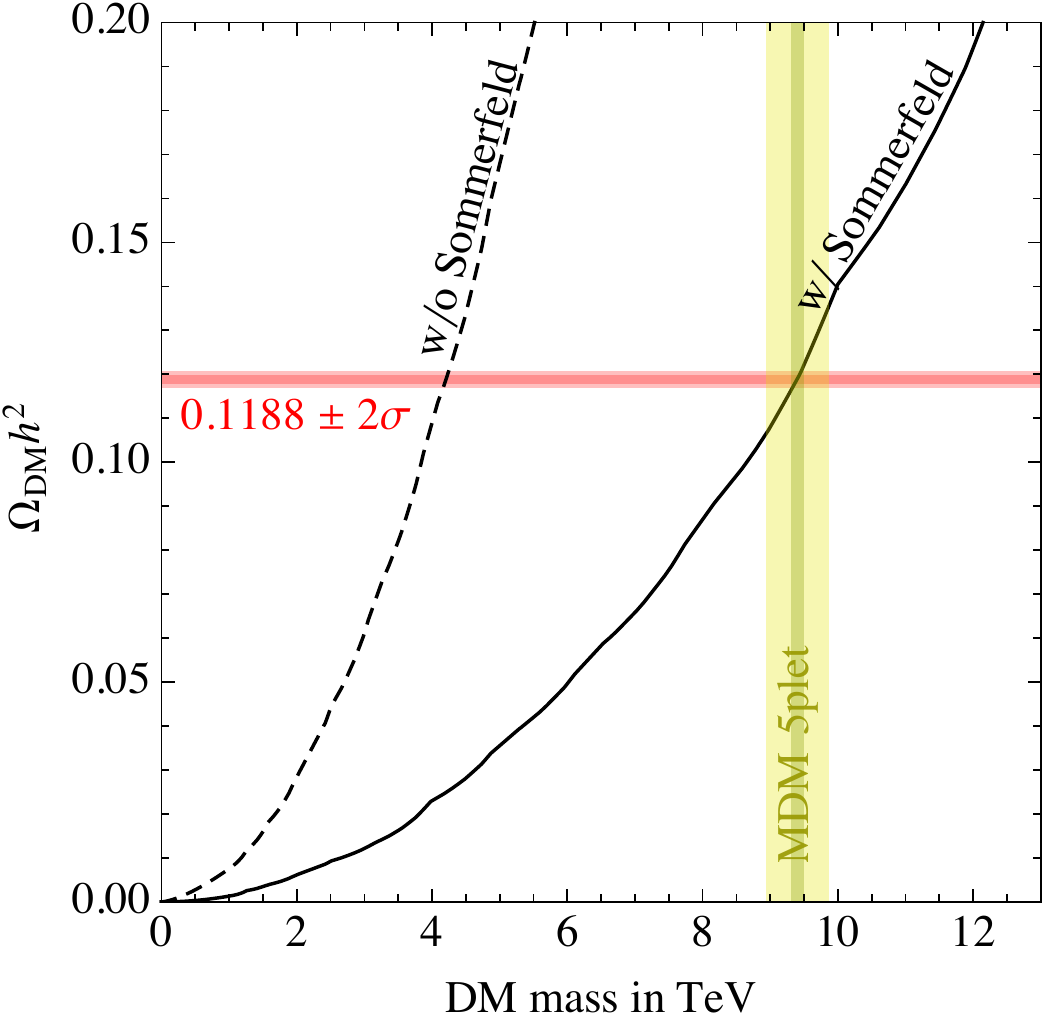} \quad
\includegraphics[width= 0.49 \textwidth]{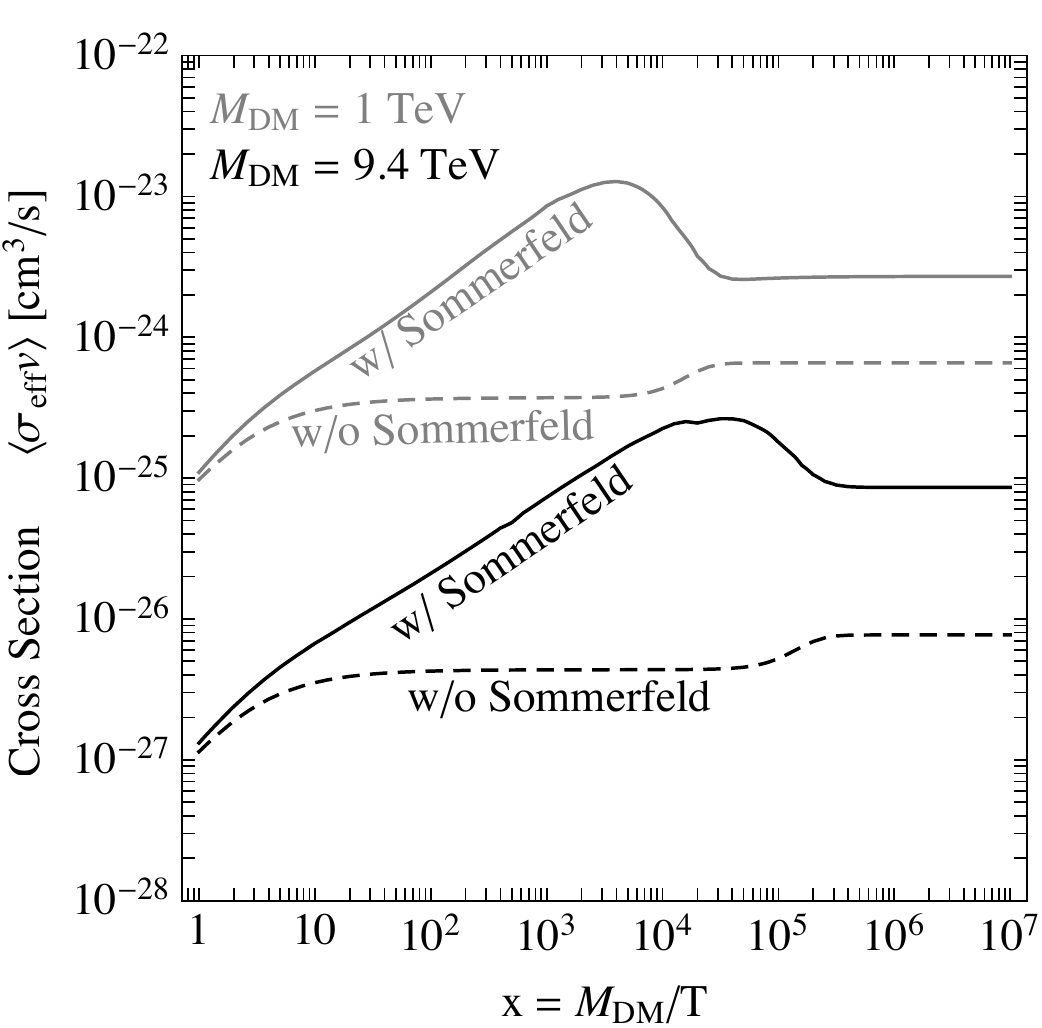}
\caption{\em \small \label{fig:relic} {\bfseries Relic abundance computation} for the MDM 5plet. Left: the DM density as a function of the DM mass, with and without the Sommerfeld enhancement. Right: the thermal averaged annihilation cross section (with and without Sommerfeld) for two indicative values of the DM mass.}
\end{center}
\end{figure}

The right panel of Fig.~\ref{fig:relic} shows the total thermally averaged annihilation cross section (with (solid lines) and without Sommerfeld (dashed lines)) for two indicative values of the DM mass (1 TeV (gray), 9.4 TeV (black)). As one can see, for $x\rightarrow 0$ (relativistic regime), $\langle \sigma_{\rm eff} v\rangle $ is not approaching a constant value because the thermal averages taken in the cosmic comoving frame and in the center-of-mass frame do not coincide in the relativistic regime \cite{Gondolo:1990dk}. As an aside the bump of the Sommerfeld enhanced cross sections, before the plateau at large $x$, is basically the convolution of the many peculiar resonant peaks of the Sommerfeld. On the other hand, the little increase at $x\gtrsim M_{\rm DM}/$(166 MeV) of the non-enhanced cross sections is due to the decoupling of the $\chi^+,\chi^-,\chi^{++},\chi^{--}$ states.   

\medskip
Having at our disposal the total thermally averaged annihilation cross section, we can finally integrate numerically Eq.~\eqref{Eq:BE} in order to determine the asymptotical value of the comoving density $Y_\infty$. After  doing so, the number density of the 5plet as a function of $M_{\rm DM}$ can be obtained via $\Omega_{\rm DM} h^2=Y_\infty M_{\rm DM} s_0 / (\rho_{c,0} h^{-2})$. Here  $s_0\simeq 2.71 \times 10^3$ cm$^{-3}$ and $\rho_{c, 0} h^{-2}\simeq 1.05 \times 10^{-5}$ GeV/cm$^3$ are the today's entropy and critical energy densities of the Universe respectively \cite{Ade:2015xua}. The left panel of Fig.~\ref{fig:relic} shows $\Omega_{\rm DM} h^2$ as a function of the DM mass. The dashed and solid lines refer respectively to the computations with and without the Sommerfeld effect, while the horizontal strips individuates the measured DM in the Universe at $1\sigma$($2\sigma$) CL by Planck 2015. The solid line crosses the 2$\sigma$ CL band when $M_{\rm DM} = 9.4 \pm 0.094$ TeV. 
We also show, as a vertical yellow band, the $\mathcal O(5\%)$ correction due to theory uncertainty on the determination of the cross sections and, in turn, on the value of $M_{\rm DM}$. In this case we get then $M_{\rm DM}=9.4 \pm 0.47$ TeV. In the rest of the paper, since the experimental error in the determination of $M_{\rm DM}$ is smaller than the theoretical one, we will always consider the latter. 

\section{Gamma ray constraints} 
\label{sec:constraints}

In this Section we compare the different components of the gamma ray spectrum, as predicted by MDM, to the experimental data. We will consider in turn the measurement of the galactic diffuse emission by \FERMI, the results on dwarf galaxies observations by, again, \FERMI, \HESS\ and \MAGIC\ and the gamma ray line searches performed by the same experiments.

\medskip

First, however, let us shortly remind of the characteristics of the DM distribution in the targets that we consider. For the Milky Way the range of possible profiles, as schematized e.g.~in~\cite{Cirelli:2010xx}, spans from the Burkert and Isothermal profiles, featuring a constant density core in the inner kiloparsecs of the Galaxy, to the peaked Navarro-Frenk-White (NFW) one, which is formally divergent as $r$ approaches the GC. The Einasto profiles are not divergent but still peaked at the GC. All these profiles need to be normalized by assuming a value for a scale radius and a scale density: we follow the procedure described in~\cite{Cirelli:2010xx} which amounts to fix the density at the location of the solar system to $\rho_\odot = 0.3$ GeV/cm$^3$ and to impose that the total DM content of the Galaxy (within 60 kpc) agrees with the recent estimates. The Einasto profiles depend on an additional parameter ($\alpha$) which controls their steepness in the GC region: as in~\cite{Cirelli:2010xx}, we will consider in some cases the standard value $\alpha = 0.17$ and also a steeper version featuring $\alpha = 0.11$ (dubbed EinastoB).

For the case of the dwarf galaxies, the uncertainty on the DM distribution reflects essentially in the uncertainty on the so-called $\bar J$ factor. We will discuss this in some detail in Sec.~\ref{sec:dwarfs}.


\subsection{Galactic gamma ray diffuse emission measurement by \FERMI}
\label{sec:diffuse}

In this Section we derive the bounds from the whole sky measurements provided by \FERMI.  
We will derive two kinds of constraints:
\begin{itemize}
\item[$\diamond$] Conservative constraints: we suppose a vanishing gamma ray background from astrophysics and we just impose that the DM signal does not exceed the measured flux.
\item[$\diamond$] Constraints obtained with a modeling of the background: this of course reduces the room for a DM signal and therefore leads to bounds that are more stringent. They are however also less robust, as they depend on the reliability of such modeling.
\end{itemize}

\subsubsection{Analysis setup}

\paragraph{Choice of the regions of interest and data extraction.}

\begin{figure}[t]
\begin{center}
\includegraphics[width= \textwidth]{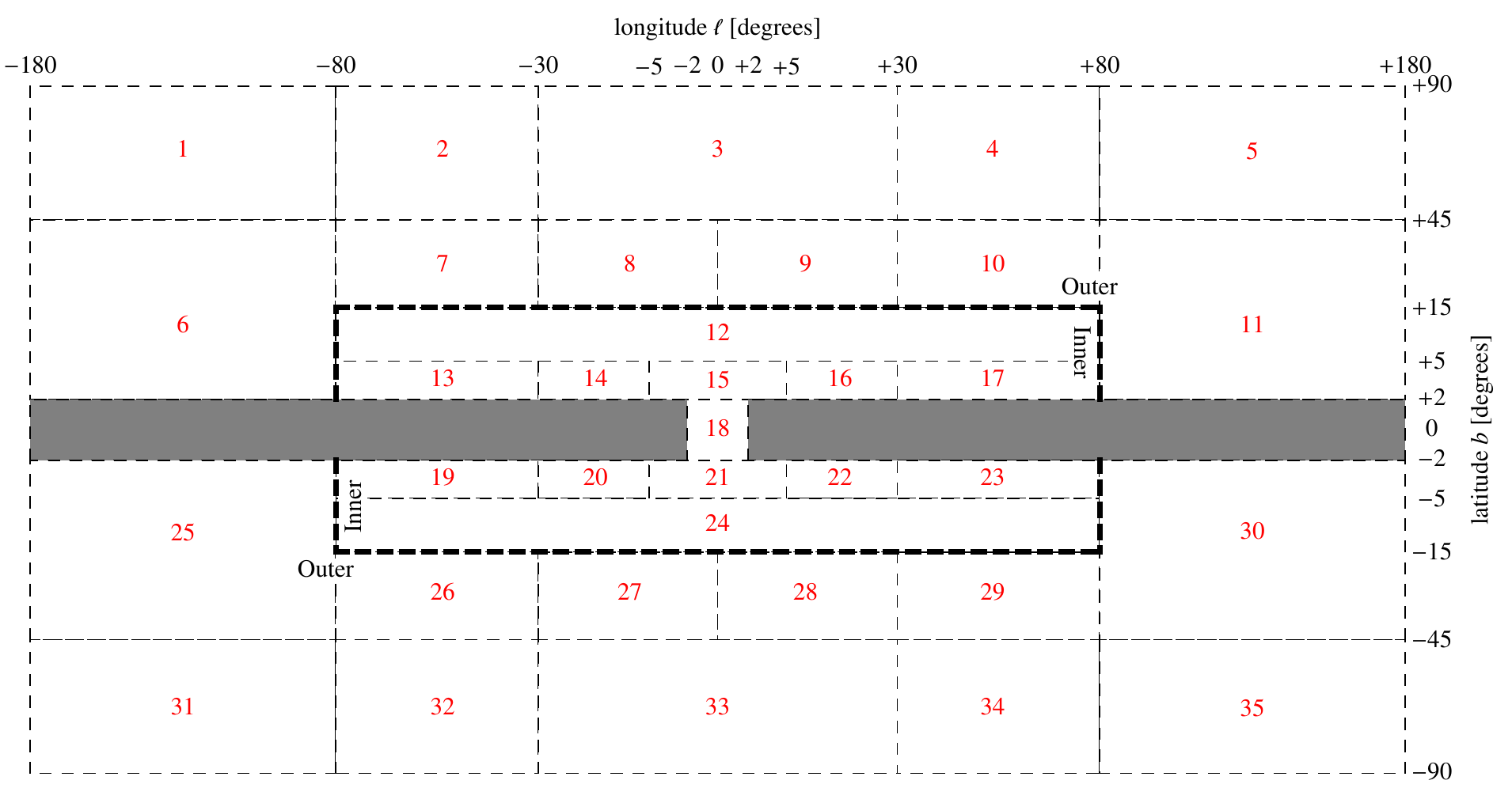}
\caption{\em \small \label{fig:galaxy_map} Slicing of the Galaxy into the different RoI's ({\bfseries Regions of Interest}). Regions from 12 to 24 included constitute our definition of Inner Galaxy. The rest constitutes the Outer Galaxy. The gray areas along the disk are masked.}
\end{center}
\end{figure}

We divide the whole galactic sky observed by \Fermi\ in 35 non-overlapping `regions of interest', as depicted in fig.~\ref{fig:galaxy_map}, masking out the $2^\circ$ around the galactic plane (but retaining a $2^\circ \times 2^\circ$ region around the GC). The regions are designed to be smaller near the GC and wider at high latitude and longitude. Some of them correspond to areas already used by previous analyses (e.g.~our RoIs number 12 and 24 coincide with the `mid-latitude strip' used by the \FERMI\ collaboration itself in~\cite{Ackermann:2012rg}), which allows us to make quantitative comparisons with previous results. For reasons that will be clear later, we distinguish between the Inner Galaxy ($|b|<15^\circ, |\ell|<80^\circ$, RoI's from 12 to 24) and the Outer Galaxy (the corresponding complement). 

\medskip

We extract \FERMI\ data using the \FERMI\ Science Tools {\tt v9r32p5}. We use $5$ years of data within the event class {\tt CLEAN}.
We perform the following selection cuts: {\tt (DATA\textunderscore QUAL==1) \&\& (LAT\textunderscore CONFIG==1) \&\& (ABS(ROCK\textunderscore ANGLE)<52) \&\& (IN\textunderscore SAA!=T)}. Events with zenith angles larger than $100^\circ$ are excluded.
The exposure is computed using the \FERMI-LAT response function {\tt P7REP\textunderscore CLEAN\textunderscore V15}.
The data are binned in 31 energy bins equally spaced in log scale between 300 MeV and 500 GeV.
Then, in order to increase the statistics at high energies, we have grouped the last four bins into two wider bins. 
The counts and the exposure maps have been produced using the {\tt HEALPix} pixelization scheme~\cite{Gorski:2004by}, with a resolution $n_{\rm side}=256$, corresponding to a pixel size of $\sim 0.23^{\circ}.$
The error bars on the differential photon flux are obtained summing in quadrature statistical and systematic (from~\cite{FermiLAT:2012aa}) uncertainties.

\paragraph{Background modeling.}
Anything other than the gamma rays produced by DM annihilation is astrophysical background for our purpose. We need therefore to have a reliable modeling of it, in order to be able to gauge the DM contribution still allowed by data. However, designing such a modeling is a challenging task by itself. We discuss in the following the procedure that we follow and how we assess its reliability. 

We consider several background components: 
\begin{enumerate}[(I)]
\item \label{diffuse} a template for the diffuse galactic emission produced by charged CR, via (\ref{diffuse}a) interactions on the interstellar gas and via (\ref{diffuse}b) the Inverse Compton process; 
\item \label{pointsources} a template for point sources, as derived in~\cite{pointsources_catalog,TheFermi-LAT:2015hja}; 
\item \label{bubbles} a template for the so-called `\FERMI\ bubbles', as provided in~\cite{Fermi-LAT:2014sfa}; 
\item \label{isotropic} the isotropic gamma ray background as measured in~\cite{Ackermann:2014usa}, including their estimate for the irreducible charged CR contamination. 
\end{enumerate}
While the components (\ref{pointsources}), (\ref{bubbles}) and (\ref{isotropic}) are rather straightforwardly implemented, (\ref{diffuse}) deserves a dedicated discussion. 
The \FERMI\ collaboration provides the template for (\ref{diffuse}) (in the supplementary material of~\cite{Ackermann:2014usa}), which we adopt. They provide three different versions (Model A, B and C, with A being the benchmark one, B adding extra CR sources and C considering variations of the charged CR diffusion coefficient): we adopt Model A for definiteness and we have checked that our procedure is only very marginally affected if choosing another one. The collaboration then corrects the template with renormalizing coefficients, energy bin per energy bin, in order to best-adapt it to the data in the wide region $|b| > 20^\circ$. We follow therefore the same procedure, by inferring the renormalization coefficients from~\cite{Ackermann:2014usa}. This is however subject to two caveats. First, we have found that the inferred coefficients are good up to a $\sim$20\% error, since in~\cite{Ackermann:2014usa} some of the subdominant foregrounds accounted for in the analysis have not been shown explicitly (\cite{fermiprivate}). Second, we are applying the renormalizing coefficients deduced from the wide $|b| > 20^\circ$ region to our RoI's which are in general different (smaller and covering also lower latitudes). However, {\it a posteriori} we will find that the agreement with the data is good within the limits that we impose (see below). Following the procedure of the collaboration (see sec.~3 of~\cite{Ackermann:2014usa}), we infer a renormalizing coefficient for each energy bin up to 13 GeV and a single coefficient for all data points above that threshold. This is done separately for the (\ref{diffuse}a) and (\ref{diffuse}b) components.

The procedure discussed above is as accurate as possible for the scope of our analysis. We test it against the data in the different RoI's and we indeed find that it provides a good description of the background for photon energies within the following windows: for the RoI's in the `inner galaxy' $1.5 \ {\rm GeV} \lesssim E_\gamma \lesssim 500$ GeV; for the `outer galaxy' $1.5 \ {\rm GeV} \lesssim E_\gamma \lesssim 100$ GeV. Indeed, the reliability of our template for point sources worsens rapidly beyond 100 GeV in the outer regions. Also, in these `low contrast' areas a more rigorous treatment of the contamination from charged cosmic rays would probably be needed.

\medskip

To summarize: we borrow the background modeling used by the \FERMI\ collaboration for their galactic diffuse study, but we adapt it to our spatial RoI's and we limit our analysis to the energy ranges in which we find that it provides a reasonable description. In doing so we will derive constraints that are necessarily less stringent, i.e.~more conservative than they could in principle be (if we had kept the whole dataset and if we had re-optimized the background to all the regions).

\paragraph{Dark Matter contribution.} 
5plet Dark Matter annihilations contribute to the galactic halo continuum gamma ray spectrum measured by \FERMI\ mainly via the DM DM $\to W^+W^-$ channel but also via the channels DM DM $\to ZZ$ and DM DM $\to Z \gamma$. To a much lesser extent, also the DM DM $\to \gamma \gamma$ contributes: EW radiation can degrade the monochromatic final state photons and generate a continuum lower-energy flux. The latter three channels give a contribution which is as relevant as the $W^+W^-$ one, thanks to the Sommerfeld enhancement of their cross sections, as discussed in sec.~\ref{sec:sommerfeld}. We therefore sum the gamma ray yields from all these processes, with the relative cross sections precisely computed in sec.~\ref{sec:sommerfeld}.

In addition, besides the gamma rays promptly emitted in the annihilation, we also include in the computation of the full spectrum the Inverse Compton Scattering secondary gamma rays (which originate when electrons and positrons from the DM annihilation scatter against the ambient galactic light). For this purpose we use the tools provided in~\cite{Cirelli:2010xx,Buch:2015iya}, to which we refer for all details.

\smallskip

It is worth noticing that, while we are here interested in considering the specific case of the 5plet, the bounds that we obtain can be safely extended to a more general class of models, in which multi-TeV DM generically annihilates into gauge bosons.
This is because the shapes of the continuum gamma rays from any $VV$ channel (with $V$ a gauge boson) are essentially indistinguishable from one another, as explicitly shown e.g.~in fig.~3 of \cite{Cirelli:2010xx}, apart from the features at the endpoint of the spectrum which are present for the $Z\gamma$ and $\gamma\gamma$ channels (from the monochromatic $\gamma$) and for the $W^+W^-$ one (from the final state radiation of a hard photon). Hence, provided that one considers DM heavy enough that these features fall beyond the sensitivity range of \FERMI\ (i.e.~for $M_{\rm DM} \gtrsim 500$ GeV), the constraints that we derive apply to the spectral shape corresponding to any gauge boson channel. One just needs to rescales the bounds with the different $\gamma$-ray multiplicities from the different channels.

\subsubsection{Background-free conservative constraints}

\begin{figure}[!t]
\begin{center}
\includegraphics[width= 0.48 \textwidth]{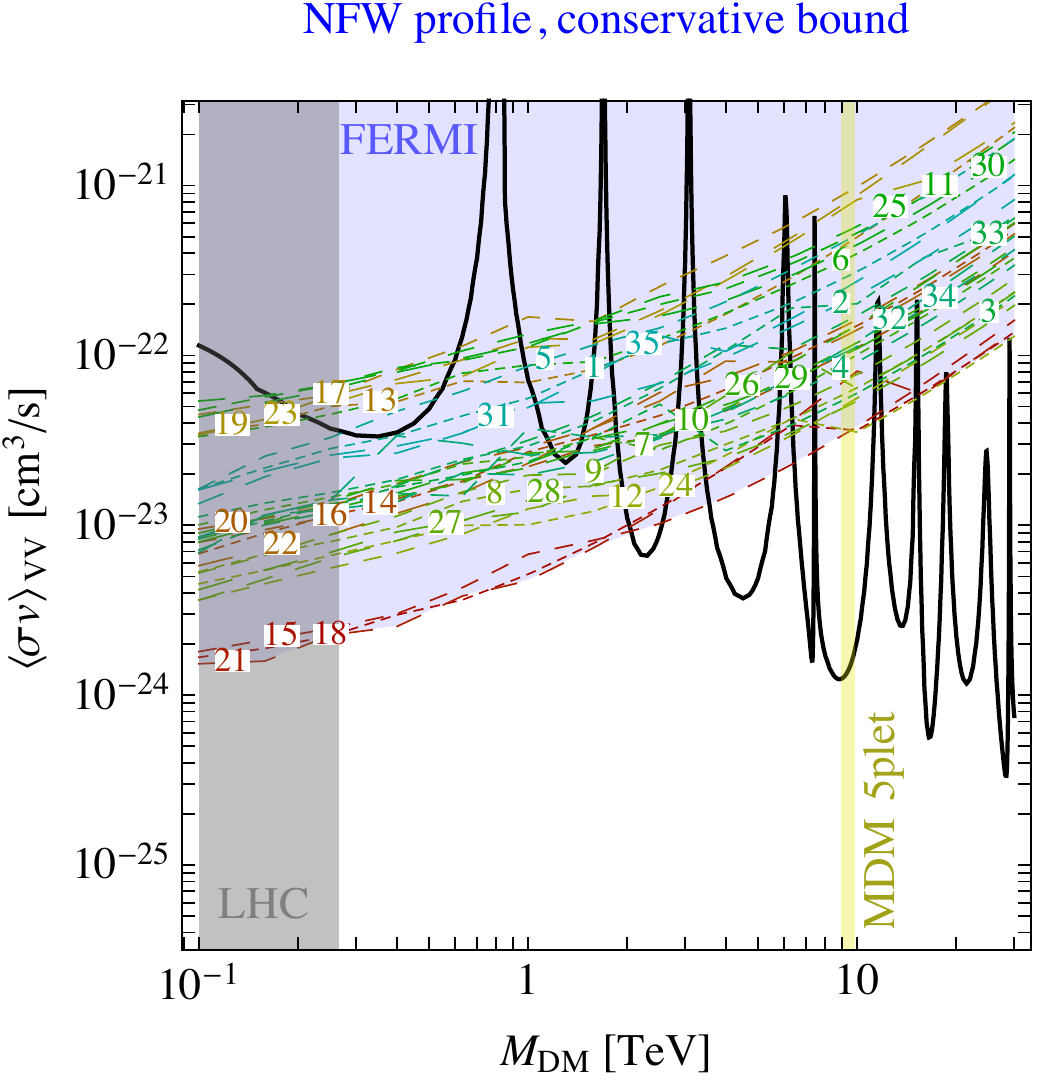} \quad
\includegraphics[width= 0.48 \textwidth]{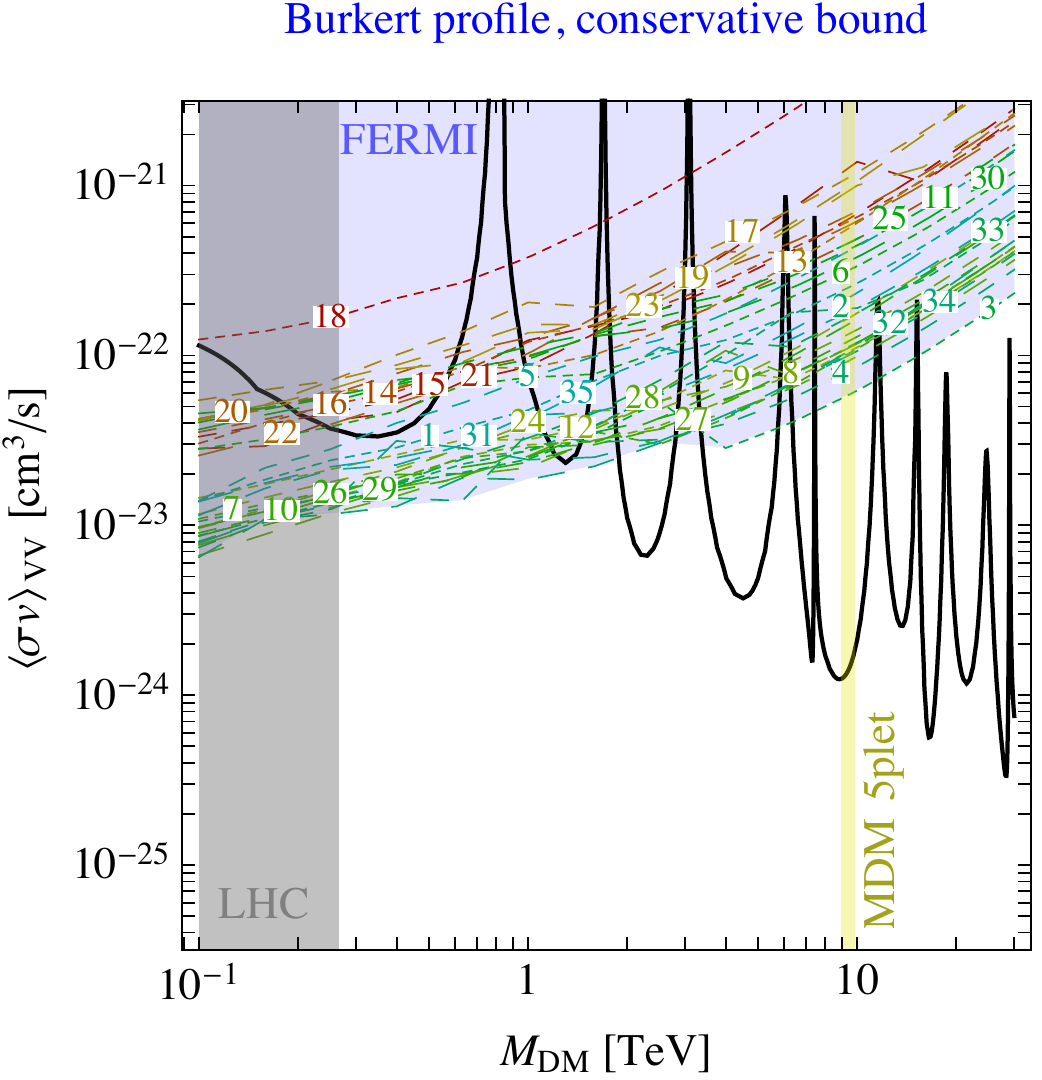}\\[6mm]
\includegraphics[width= 0.48 \textwidth]{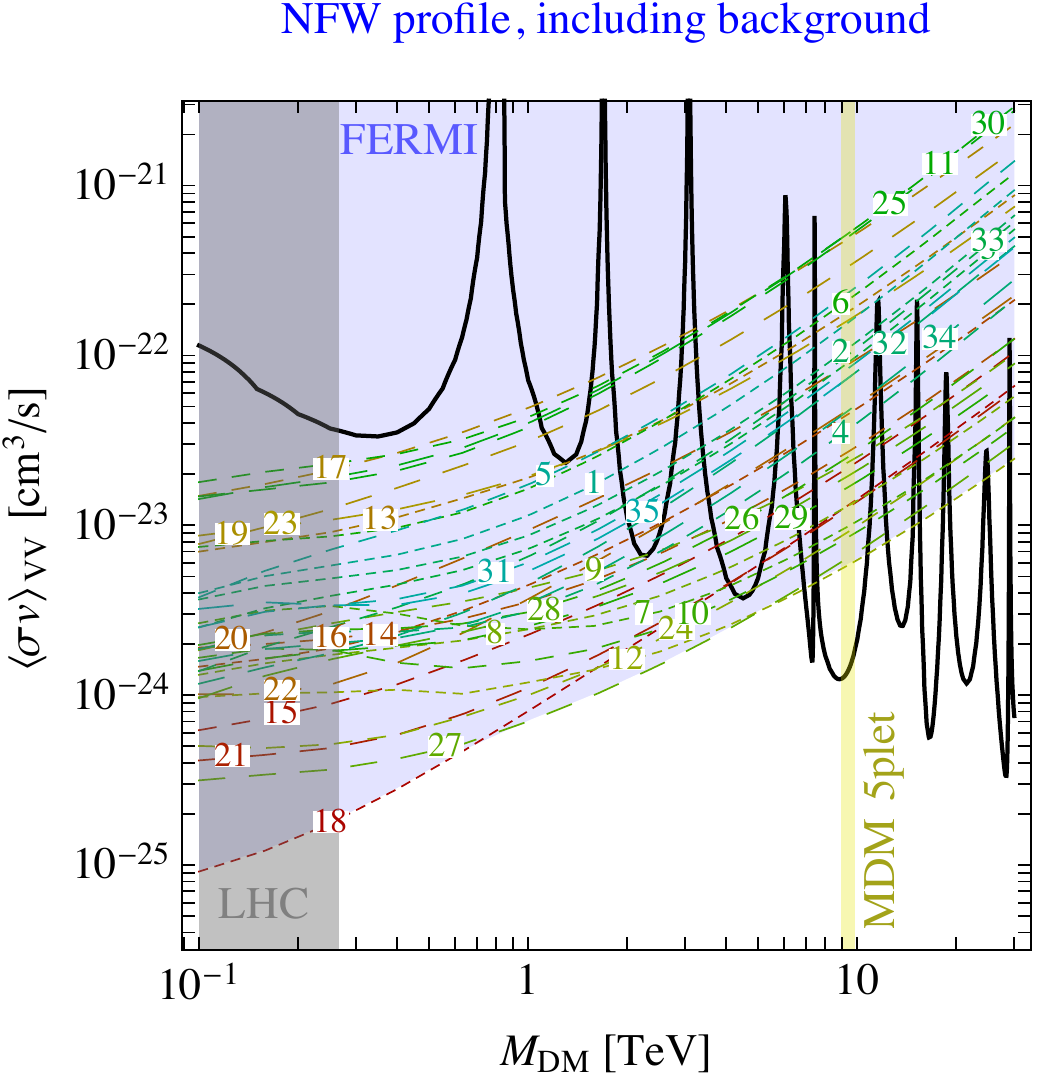} \quad
\includegraphics[width= 0.48 \textwidth]{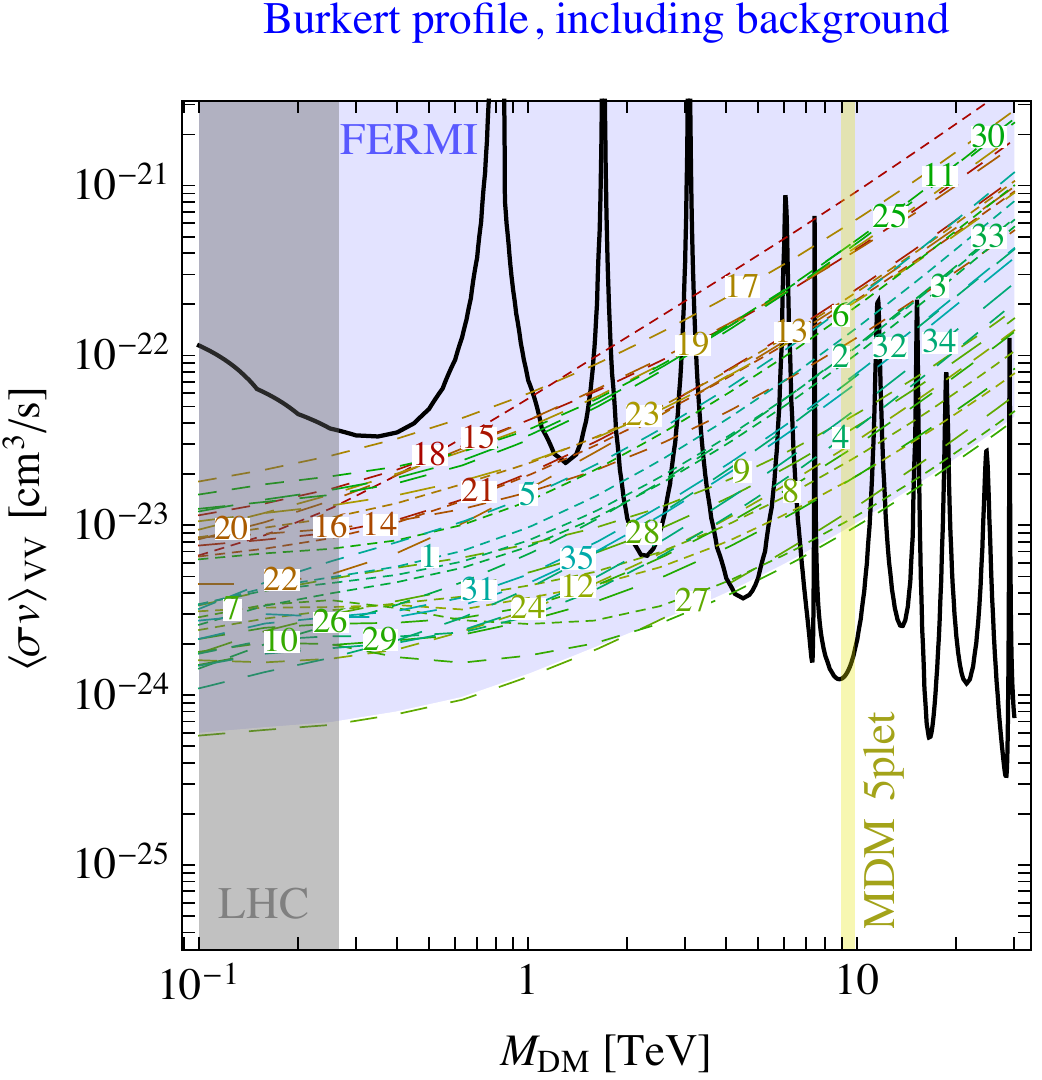}
\caption{\em \small \label{fig:bounds_continuum} {\bfseries Constraints from diffuse \FERMI\ data}. Each line corresponds to one of the Regions of Interest numbered in fig.~\ref{fig:galaxy_map}: the area above the curve is excluded. Left: NFW DM profile; right: Burkert profile. Top: conservative constraints without background; bottom: constraints including background. The left portion of the planes is excluded by the LHC constraints obtained in~\cite{Ostdiek:2015aga}.}
\end{center}
\end{figure}

\begin{figure}[t]
\begin{center}
\includegraphics[width= 0.48 \textwidth]{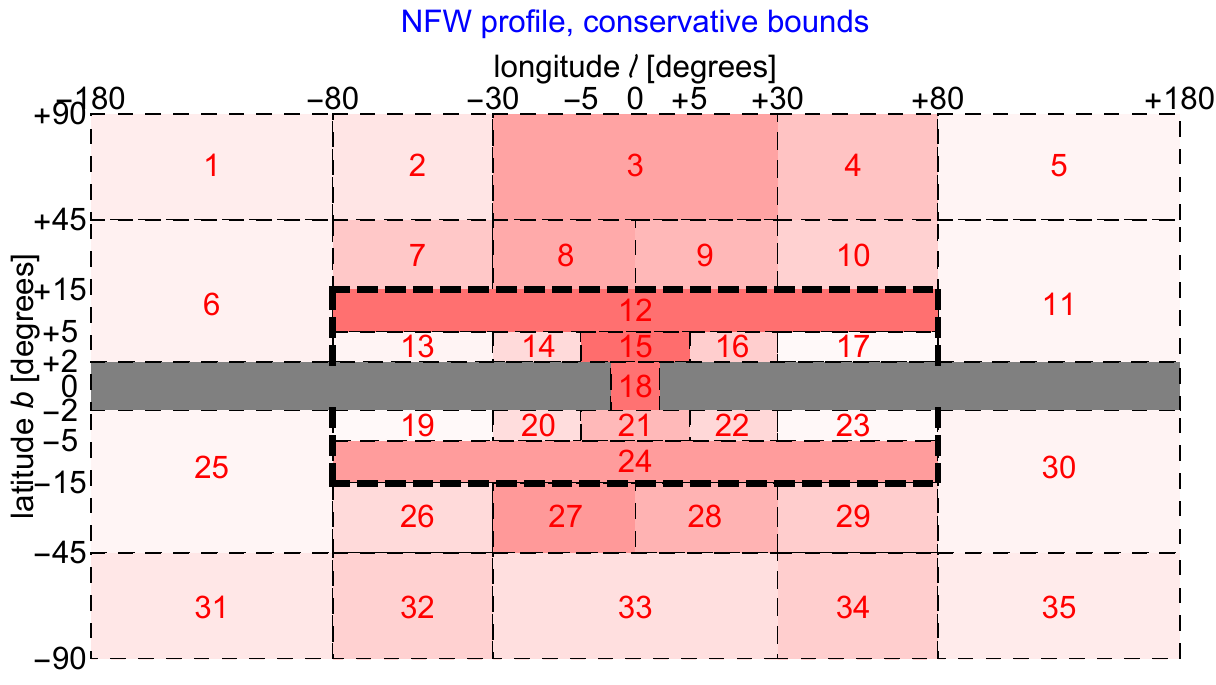} \quad
\includegraphics[width= 0.48 \textwidth]{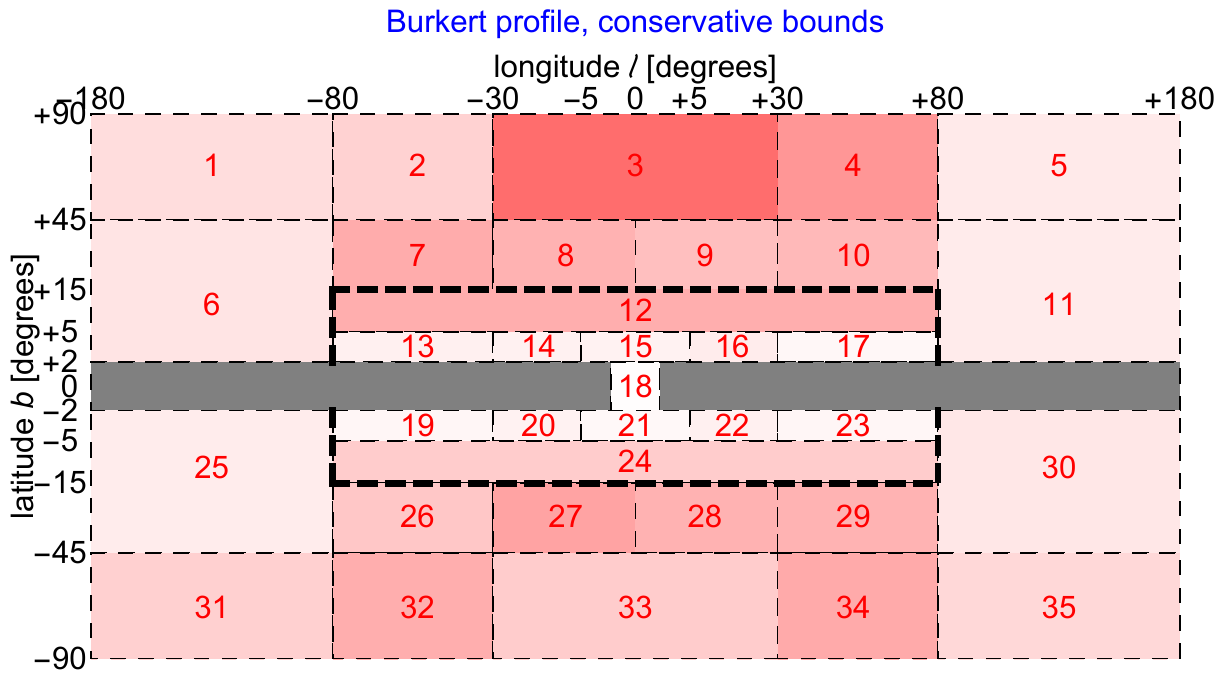}\\[3mm]
\includegraphics[width= 0.48 \textwidth]{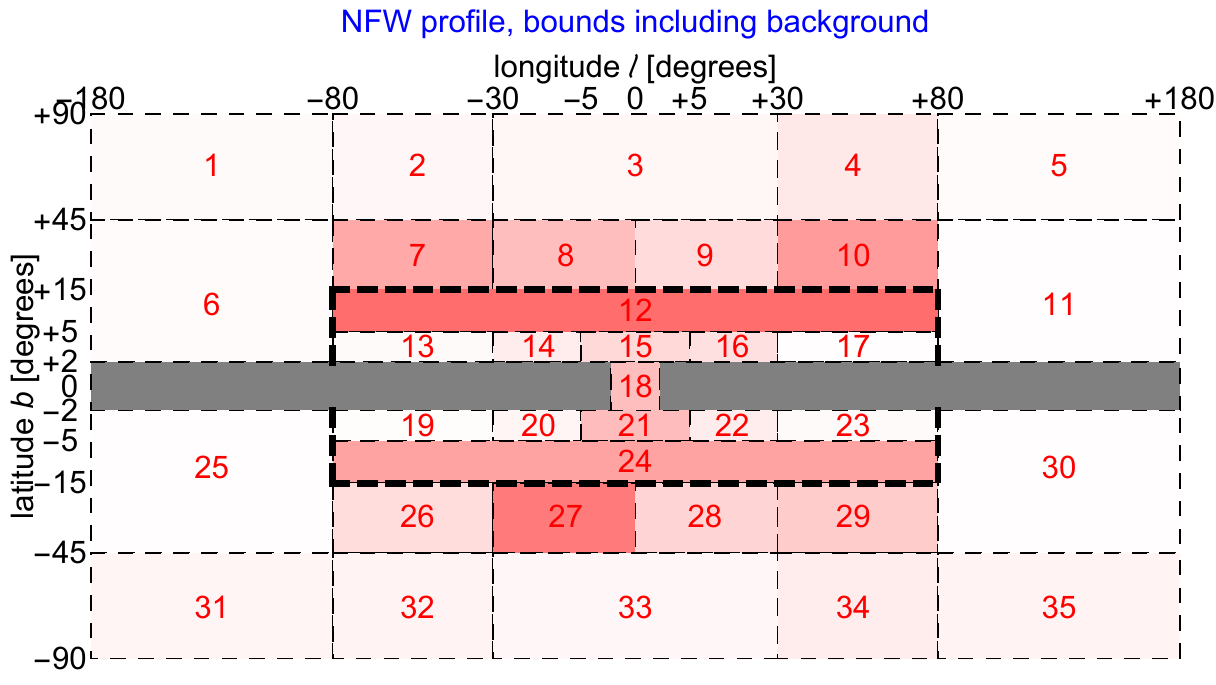} \quad
\includegraphics[width= 0.48 \textwidth]{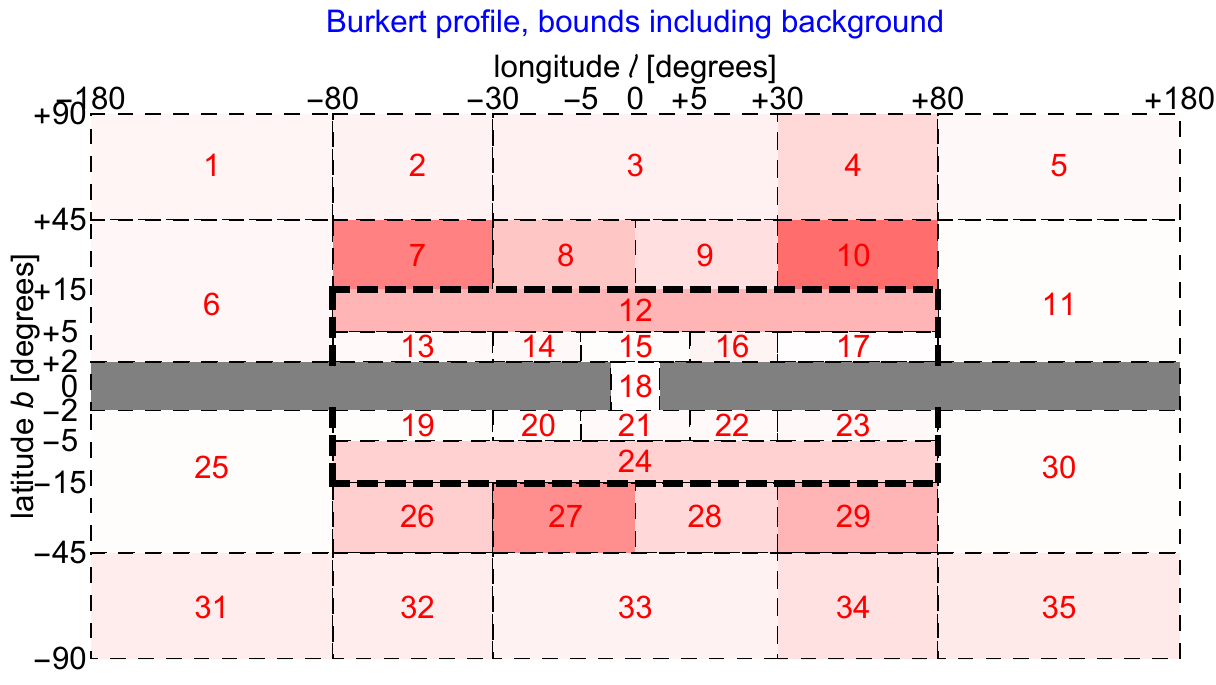}
\caption{\em \small \label{fig:bounds_continuum_map} Same as in fig.~\ref{fig:bounds_continuum}, in an alternative visualization of which regions impose the most stringent constraints, for the specific value $M_{\rm DM} = 9.4$ TeV (the thermal mass of the MDM 5plet). The redder the shading, the stronger the constraint (we use a different normalization of the color scale per each chart).}
\end{center}
\end{figure}

In fig.~\ref{fig:bounds_continuum} (upper panels) we present the constraints that we obtain by imposing that the DM signal does not exceed by more than $3\sigma$ any of the data points in any given RoI. We consider two representative DM profiles (NFW and Burkert) and we span a large range of DM masses, from 100 GeV to 30 TeV. 

We see that there is no single RoI which dominates over all the range of masses. However, as expected, regions close to the GC (e.g.~RoI 21, 15 and 18) impose the strongest constraints for the peaked NFW profile while wide regions in the outer galaxy (e.g.~3, 4, 27) are the most relevant for the cored Burkert profile. To better illustrate this point, in fig.~\ref{fig:bounds_continuum_map} we choose a definite value of the DM mass (9.4 TeV, as obtained in sec.~\ref{sec:relic}) and we report on the galaxy map the strength of the different constraints. 

On the other hand, considering the envelop of the most constraining curves on the whole range of $M_{\rm DM}$, we see that the NFW hypothesis provides  stronger bounds than those from Burkert, but the difference never goes beyond a factor of a few.

\medskip

In fig.~\ref{fig:bounds_continuum} we superimpose the total cross section in the channels mentioned above~\footnote{We denote the bounds as being on the annihilation cross section DM DM $\to VV$, with $V$ a gauge boson $W, Z$ or $\gamma$. Namely, here we have computed the bounds considering the case of a pure $W^+W^-$ annihilation, but they can be recast for different channels. We refer to the discussion in the previous paragraph for more details.}. The mass intervals for which the line enters in the shaded regions are excluded. We see that, while significant constraints can be imposed especially below $\sim$2 TeV, the model is still allowed for large intervals of masses. For the specific case of {\em the} MDM 5plet ($M_{\rm DM} \simeq 9.4$ TeV), the bound lies almost 2 orders of magnitude above the predicted cross section, thanks to the fact that the Sommerfeld enhancement displays a trough at that value in mass.

\subsubsection{Constraints including background}

In fig.~\ref{fig:bounds_continuum} (lower panels) we present instead the constraints that we obtain by including an astrophysical background. Here the bounds are derived by looking for the best fit (background + DM) configuration and then requiring that the addition of more DM does not worsen the best fit $\chi^2$ by more than $\Delta \chi^2 = 9$. The situation changes in two respects. First, of course the bounds are much stronger, as less room is left for Dark Matter. Second, the relative importance of the different RoI's in setting the bounds changes, as a consequence of the fact that our background modeling may describe accurately or not the measured flux in any specific RoI. Again in fig.~\ref{fig:bounds_continuum_map} we visualize the most important regions from which the constraints originate. 

\medskip

We see that now the measurements rule out essentially all the region below $M_{\rm DM} \lesssim 7$ TeV. For larger masses, small islands are excluded up to about 25 TeV. {\em The} MDM 5plet is again spared both for the NFW and the Burkert cases. The constraints from NFW and from Burkert are now even more similar.


\subsection{Dwarf galaxies observations of gamma ray continuum}
\label{sec:dwarfs}

Dwarf satellite galaxies are believed to be some of the cleanest possible laboratories to search for DM in gamma rays, thanks to their presumed high DM content and relatively reduced stellar emission foreground. On the other hand, the scarcity of stellar tracers makes it difficult to precisely reconstruct how much DM they actually host and how it is distributed, leading to large uncertainties. For the case at hand, a related uncertainty has to do with the Sommerfeld enhancement: in order to compute it precisely, one would of course need to know the DM velocity dispersion, which is in principle different in each galaxy and in different radial positions within each galaxy. As customary, we assume that the DM velocity is the same as that of the stellar tracers (which is plausible if the systems have reached relaxation) and we adopt a common value of $10 \ {\rm km/s} = 3 \times 10^{-5}\, c$, in line with the measurements that typically span 3 to 15 km/s (see e.g.~\cite{Bonnivard:2015xpq} for a recent compilation).

The uncertainties on the DM content and distribution for each galaxy are the subject of a long ongoing debate in the literature (see e.g.~\cite{Evans:2003sc,Strigari:2007at,Essig:2009jx,Charbonnier:2011ft,GeringerSameth:2011iw,Martinez:2013els}). They are commonly expressed as uncertainties in the determination of the so-called average $\bar J$ factor,
\begin{equation}
\bar J = \frac{1}{\Delta \Omega} \int_{\Delta \Omega} d\Omega\int_{\rm l.o.s.} d s \, \rho(s,\theta)^2,
\end{equation}
where $\Delta \Omega$ is angular area subtended by the dwarf spheroidal galaxy or by the instrument 
and $\theta$ is the angle between the axis connecting the Earth and to the dwarf galaxy and the line of sight.
The \FERMI\ collaboration quotes a rather small uncertainty at the level of 10\% to 40\% at most~\cite{Ackermann:2015zua} for the stacked sample of galaxies. For the individual dwarf galaxies (in the `classical' class), a recent study~\cite{Geringer-Sameth:2014yza} finds values from 17\% to 124\%.
The very recent dedicated study in~\cite{Bonnivard:2015xpq} typically finds larger uncertainties, which can however differ case by case. Finally, in case of somewhat more exotic scenarios in which dwarf galaxies host intermediate mass black holes, the flux can be modified by a factor from a few up to $10^6$~\cite{Gonzalez-Morales:2014eaa}.
In view of this situation we are prompted to consider, alongside our constraints, alternative ones that are somewhat relaxed (as we will detail below). Future observational work on these systems has clearly the potential of reducing these uncertainties significantly. 

\medskip

We base our analysis mainly on the most recent observation of 15 dwarf galaxies by \FERMI~\cite{Ackermann:2015zua}. We also consider the constraints imposed by \HESS~\cite{Abramowski:2014tra}, which has observed a subset of four of these dwarf galaxies, plus Sagittarius (not included in the \FERMI\ analysis), and \MAGIC~\cite{Aleksic:2013xea}, which has intensively observed Segue1. The \HESS\ constraints complement those of \FERMI\ in the window $M_{\rm DM} = 10 \to 20$ TeV while the \MAGIC\ ones remain subdominant to those of \FERMI, since they do not extend beyond 10 TeV.

\begin{figure}[t]
\begin{center}
\includegraphics[width= 0.48 \textwidth]{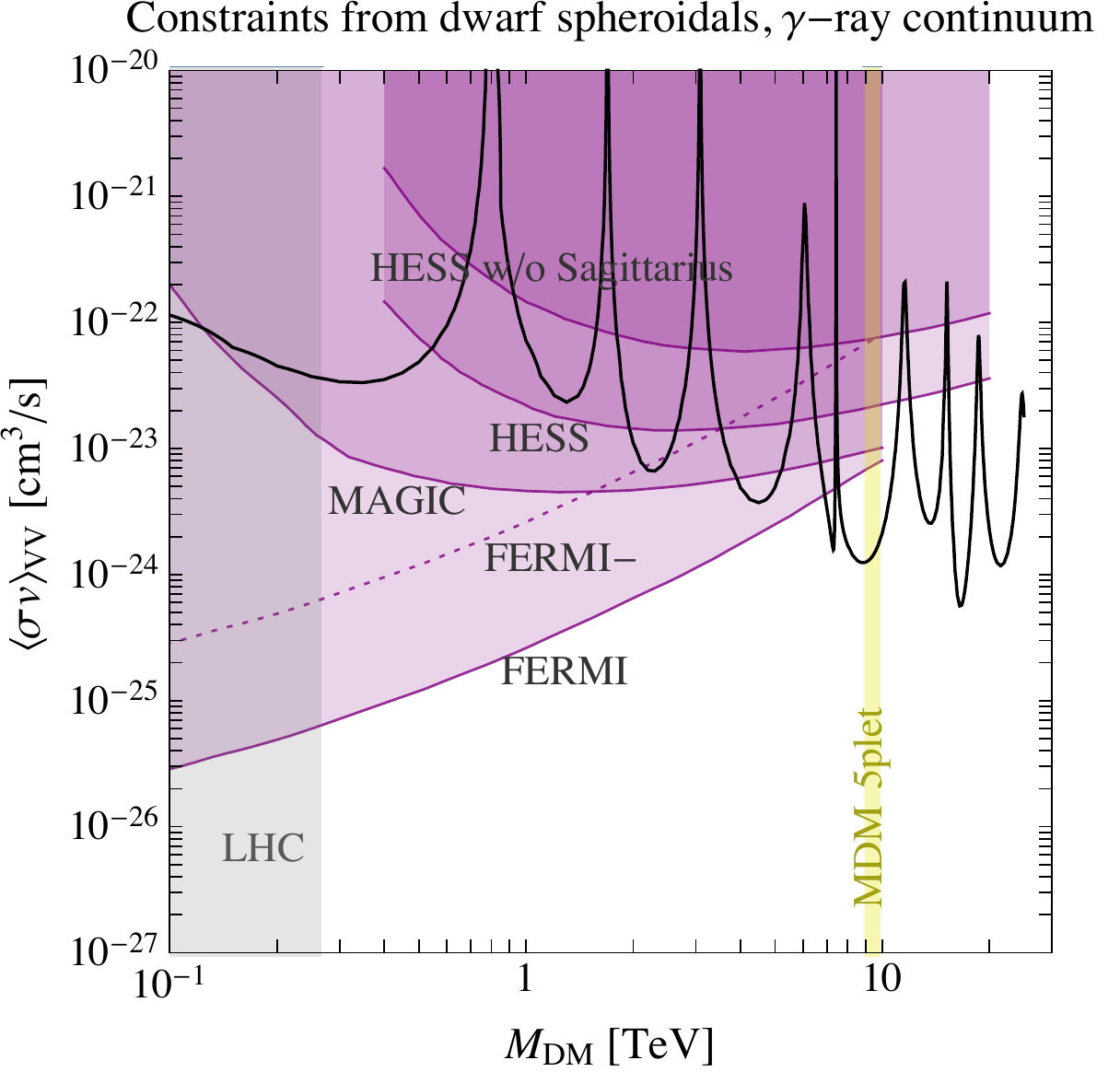} 
\caption{\em \small \label{fig:dwarves} {\bfseries Constraints from dwarf spheroidal galaxies} (solid colored lines and shaded areas) superimposed to the predicted DM cross section into $VV$ (black line). The dotted line labelled `\FERMI -' corresponds to a conservative estimate of the uncertainty associated to $\bar J$ factors in dwarves (see text).}
\end{center}
\end{figure}

In fig.~\ref{fig:dwarves} we report the bounds on the annihilation in the $W^+W^-$ annihilation channel obtained by the experimental collaborations and we compare them with the predicted 5plet DM cross section into $W^+W^-+ZZ+Z\gamma/2$, which we denote as $VV$ like before. Indeed, the spectral shapes of {\em continuum} $\gamma$-rays from these channels are very similar (as discussed above) and we neglect here the largely subdominant continuum $\gamma$-rays from the $\gamma\gamma$ channel. The fact that the $W^+W^-$ channel features a sharp peak at $E \sim M_{\rm DM}$, from the emission of a hard final state photon, is not expected to modify these results significantly, also in light of the much larger uncertainties connected to the DM contents determination.

We find that the constraints from \FERMI, taken at face value, exclude all the range $M_{\rm DM} \lesssim 7$ TeV. We also consider bounds relaxed by an order of magnitude (labelled `\FERMI -'), intended as an average conservative assessment of the uncertainties related to $\bar J$-factors discussed above. This slightly reduces the excluded interval to $M_{\rm DM} \lesssim 6.3$ TeV. 
The constraints from \HESS\ rule out very small intervals between 10 and 20 TeV. The bounds that do not include the most stringent dwarf (Sagittarius) are slightly less constraining. 

We stress that all these numbers are strongly dependent on the detailed shape and position of the Sommerfeld peaks, which in turn suffer from a `theory uncertainty' that we have quantified before to be of 5\% to 10\%. However, the global results are robust.
The mass for {\em the} MDM 5plet is again not affected by these bounds, unless very aggressive assumptions on the $\bar J$-factor determinations, such as  e.g.~to lower significantly the \FERMI\ bound, are made.

\medskip

To conclude this section, we point out that significant progress could be made by current and future experiments. Improving the \FERMI, \MAGIC\ or \HESS\ bounds by a factor of just a few at the largest masses would allow to probe almost the entire parameter space of the model. In this respect, choosing one of the dwarf galaxies with the most promising $\bar J$-factors (such as Coma or Ursa Major II, according to~\cite{Bonnivard:2015xpq}) can perhaps allow to reach such a goal.


\subsection{Gamma ray lines searches}
\label{sec:lines}

\begin{figure}[t]
\begin{center}
\includegraphics[width= 0.48 \textwidth]{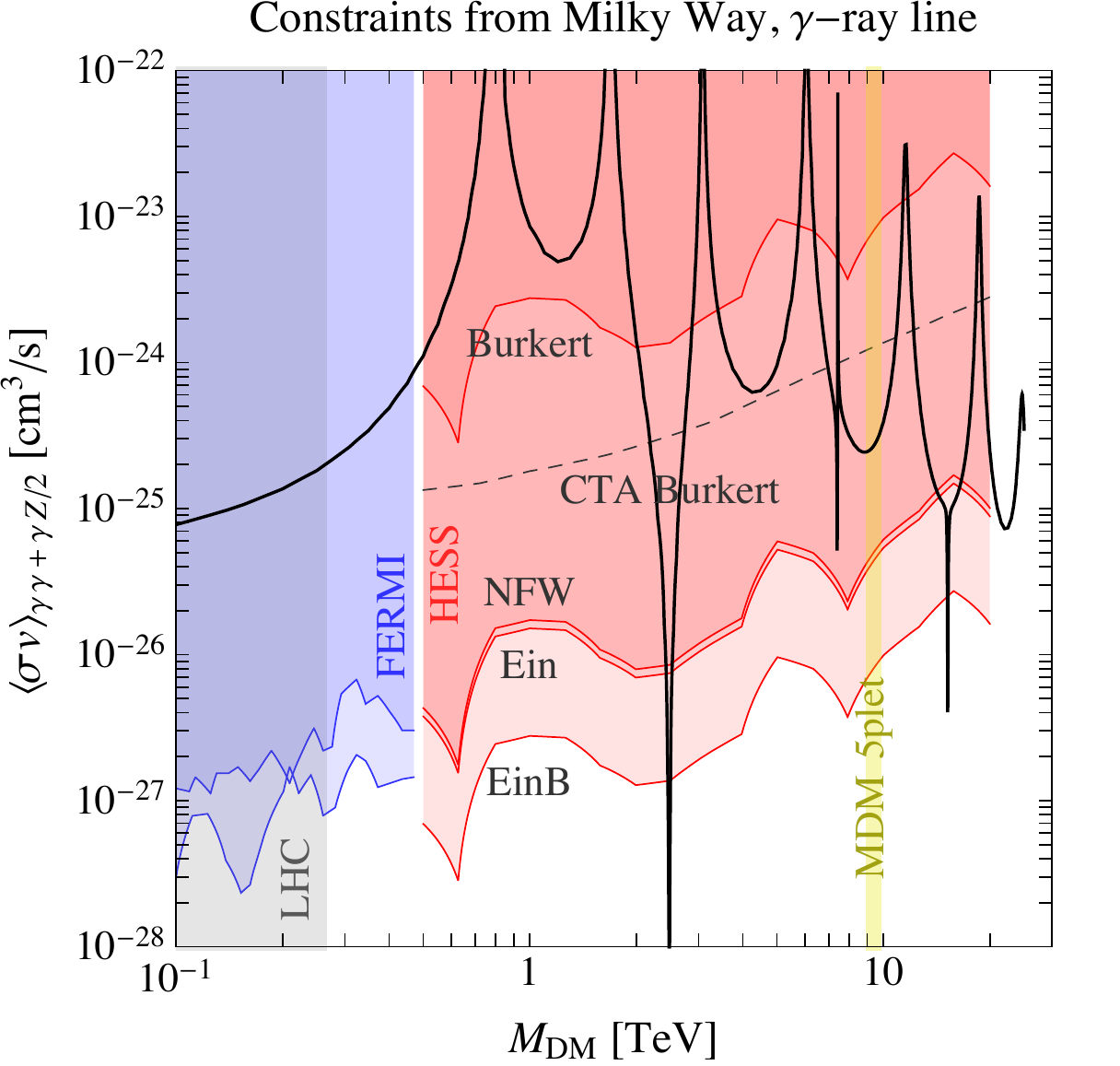} \quad
\includegraphics[width= 0.48 \textwidth]{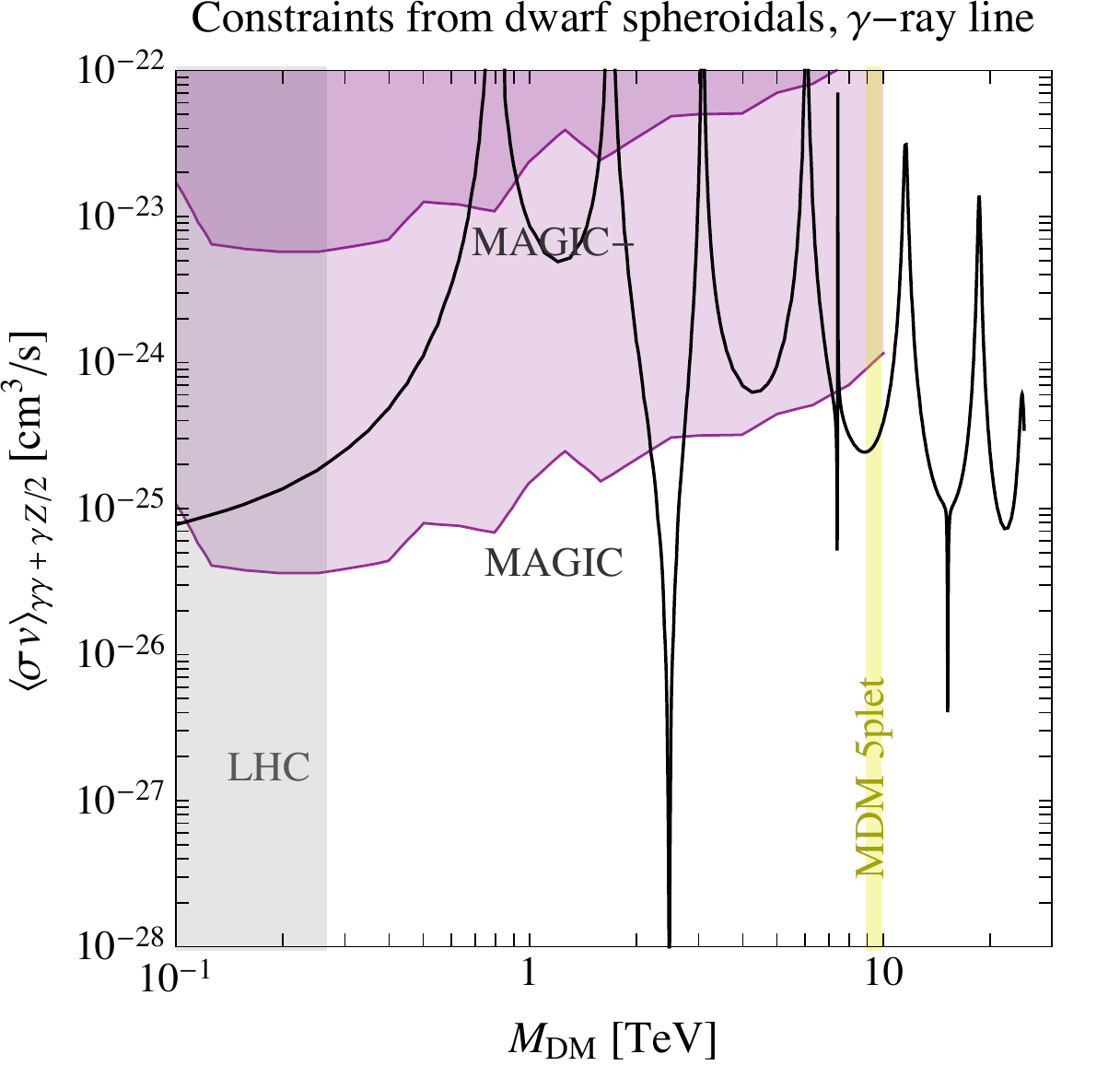}
\caption{\em \small \label{fig:lines} {\bfseries Constraints from gamma ray line searches}: in the GC region by \HESS\, and \FERMI\, (left panel) and in the dwarf spheroidal galaxy Segue1 by \MAGIC\, (right panel).}
\end{center}
\end{figure}

As already mentioned above, $\gamma$-ray lines (or sharp spectral features which are degenerate with lines for experimental purposes) arise at the endpoint of the spectrum from DM annihilation. Searching for these features has been regarded since a long time as a very promising strategy, and they have been often referred to as the proverbial `smoking gun' for Dark Matter~\cite{Bergstrom:1988fp,Rudaz:1989ij,Bouquet:1989sr,Giudice:1989kc,Rudaz:1990rt,Bergstrom:1997fj,Weniger:2012tx,GeringerSameth:2012sr,Gustafsson:2013gca,Tempel:2012ey,Chu:2012qy,Bergstrom:1989jr,Bringmann:2007nk,Ibarra:2012dw}.

\medskip

The \HESS\ telescope has imposed line constraints in~\cite{Abramowski:2013ax}, the relevance of which, for pure wino DM, was initially recognized by~\cite{Cohen:2013ama,Fan:2013faa}. 
The region observed by \HESS, the Central Galactic Halo (CGH), is a promising one due to its relative proximity and large predicted DM concentration. The search region is defined as a circle of $1^\circ$ radius centered on the GC, where the Galactic plane is excluded, by requiring $|b| > 0.3^\circ$.~\footnote{The \HESS\ search is performed adding to the background a Gaussian-shaped line, in different bins of energy, and thus is effective for flat DM profiles. This is opposite to the \HESS\ searches for a $\gamma$ ray signal in the inner region of the galaxy~\cite{Abramowski:2011hc}, whose constraints do not apply for profiles with a large core, since there the background is estimated with the On/Off procedure from a control region close to the center.}

\MAGIC\ has also published $\gamma$-ray line constraints~\cite{Aleksic:2013xea} from the observation of the Segue 1 dwarf galaxy. Finally, we will also quote the bounds from the \FERMI\ collaboration, presented in~\cite{TheFermi-LAT:2015gja}, which are relevant at small DM masses.

\medskip

We consider the 5plet DM annihilation channels $\gamma\gamma$ and $\gamma Z$. 
Some recent works have advocated the need to go to higher orders in the cross sections for heavy EW multiplets, and have provided refined fixed order~\cite{Hryczuk:2011vi} and resummed~\cite{Bauer:2014ula,Ovanesyan:2014fwa} calculations, in particular for the exclusive cross section into $\gamma \gamma$. The effect of these resummations is a reduction of the annihilation cross section by a factor of $\sim 3$ for EW triplets of mass of a few TeV.
Recently, however, the issue has been revisited in~\cite{Baumgart:2014saa}, where it has been pointed out that a less exclusive cross section would be more appropriate for this kind of channels, given the limited resolution (hundreds of GeV) of the current experiments for high energy $\gamma$ rays. Indeed the relevant cross section is the one for $\gamma\gamma + X$, where $X$ is anything soft enough not to affect the line, once the experimental resolution is taken into account.~\footnote{ 
We are consistently neglecting the radiation of a very hard $\gamma$ from $W$ bosons. Such an emission is Sudakov enhanced, and it gives a feature at $E_\gamma \simeq M_{\rm DM}$ which resembles a line --except for a shift of order $m_W$, which serves as a cut-off of the soft divergence--. Given that the experimental resolution is larger than $m_W$, this contribution is observationally indistinguishable from $\gamma \gamma$ and $\gamma Z$ lines. Moreover, to be consistent in including this effect, one would need to compute the same processes, at the same order in the EW couplings, including splittings like $\gamma \to WW$. For inclusive enough observables (as it is the case, given the experimental resolutions), these effects compensate with the extra photons radiated by the $W$ bosons.
} Ref.~\cite{Baumgart:2014saa} presents a result of this procedure for a pure Wino Dark Matter candidate and shows that the resummed cross section is barely distinguishable from the lowest order one, in the interesting mass range.
In light of the above discussion, as well as for simplicity, we stick to the lowest order cross section in $\gamma\gamma + \gamma Z/2$, to which we add the effect of Sommerfeld enhancement.

\medskip

In fig.~\ref{fig:lines} we show the constraints superimposed to the DM annihilation cross section as computed in sec.~\ref{sec:sommerfeld}.
Notice that \HESS\ quotes in~\cite{Abramowski:2013ax} a bound on an Einasto profile with $\alpha = 0.17$, scale radius $r_s = 20$ kpc and scale density $\rho_s = 2.8 \times 10^6 \ M_\odot/{\rm kpc}^3$,~\footnote{It is important to realize that these parameters imply a total DM content of the Milky Way and a local DM density much higher than the ones that we adopt (see the discussion at the beginning of Sec.~\ref{sec:constraints}).} which we have to rescale to our standard profiles. We see that the \FERMI\ bounds (which we report for two different profiles and regions, marginally different) rule out the low mass portion. The \HESS\ constraints are very relevant in the 500 GeV $\to$ 20 TeV window: if the profile is assumed to be peaked (such as NFW or Einasto) the whole region is essentially ruled out. If the profile is cored (such as Burkert), some intervals are reopened. 

{\em The} MDM 5plet falls in one of these intervals: if the profile is peaked, it is ruled out; if the profile is cored, it is again spared. 

The constraints from Segue1 by \MAGIC\ are equally relevant and have the power to rule out essentially the entire window up to $M_{\rm DM} = 7$ TeV. However, as we commented in Sec.~\ref{sec:dwarfs}, they are subject to a large indetermination. In particular, the Segue1 dwarf spheroidal is found in~\cite{Bonnivard:2015xpq} to have a large $\bar J$-factor uncertainty. If, for illustration, we rescale the bound by the `1$\sigma$' of the value determined in~\cite{Bonnivard:2015xpq} for the $\bar J$-factor of Segue1, the constraints relax significantly (see the line labelled `\MAGIC-' in fig.~\ref{fig:lines}).

\medskip

We conclude this section by briefly mentioning the prospects for improvements. In fig.~\ref{fig:lines} (left) we report the sensitivity reach for the {\sc Cta} observatory as quoted in~\cite{Ovanesyan:2014fwa} (on the basis of~\cite{Bergstrom:2012vd}), rescaled to the case of the Burkert profile. We see that significant additional portions of the parameter space would be covered, but still large gaps would remain at high masses, due to the peculiar peak structure of the Sommerfeld enhanced cross section. In particular, {\em the} MDM 5plet would still not be tested. 

Concerning the line searches in dwarf spheroidals, we stress again that improving the existing bounds by a relatively small factor, e.g.~focussing on target galaxies with potentially high $\bar J$-factors, would allow to complete the coverage in mass, including the value of {\em the} MDM 5plet.


\section{Conclusions}
\label{sec:conclusions}

\begin{figure}[t]
\begin{center}
\includegraphics[width= 0.8 \textwidth]{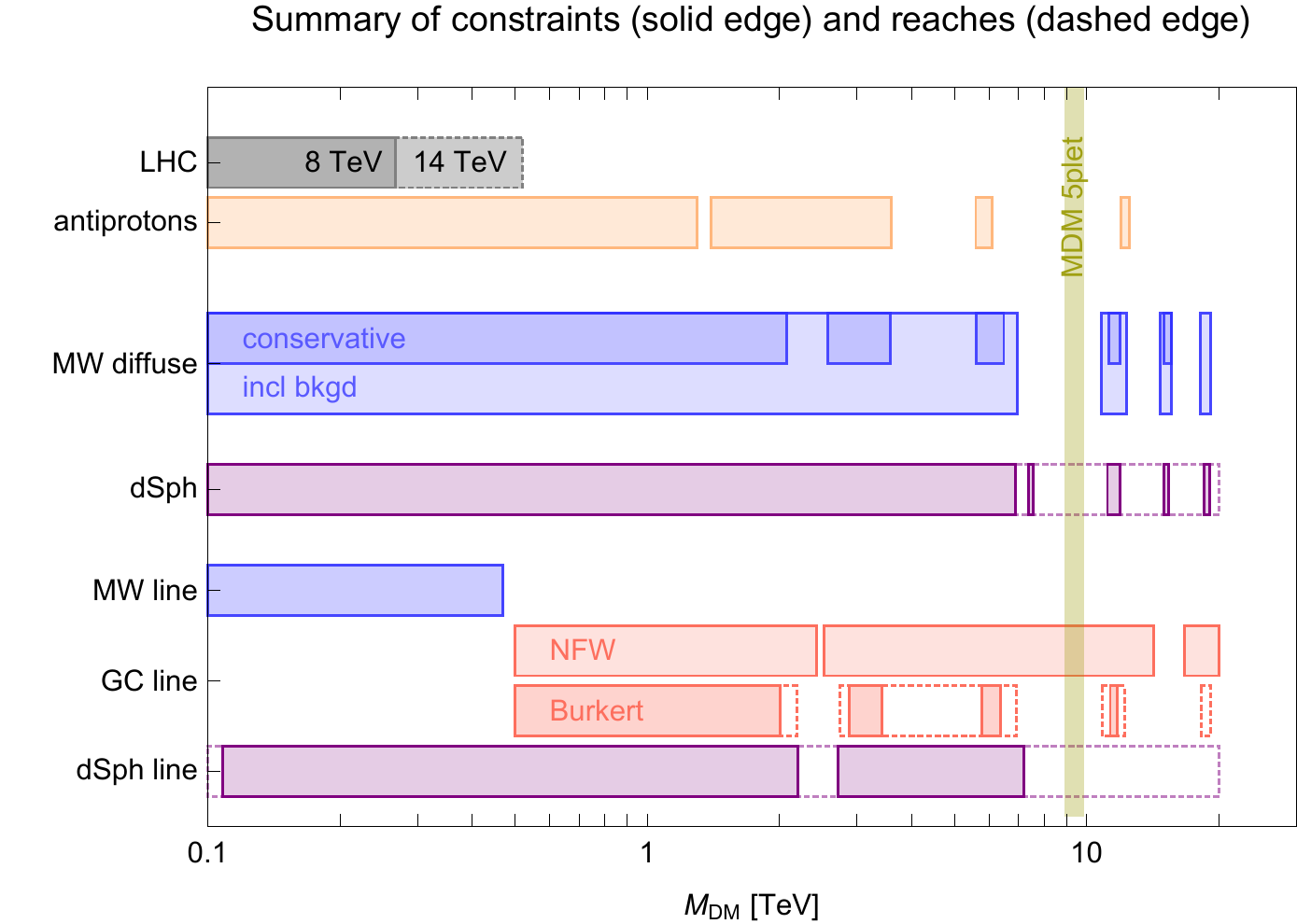} 
\caption{\em \small \label{fig:summary} {\bfseries Summary chart} of the constraints from gamma rays on 5plet Dark Matter. The vertical band individuates {\em the} Minimal Dark Matter candidate with mass fixed by thermal production.}
\end{center}
\end{figure}

Motivated by the Minimal Dark Matter model, and more generally by the model-class of multi-TeV Dark Matter with EW interactions (multi-TeV WIMPs), we have explored the constraints that come from several gamma ray probes. A crucial ingredient for these kinds of models, as recognized since quite some time, is the Sommerfeld enhancement which arises from the exchange of EW bosons among the heavy DM particles: it modifies significantly the annihilation cross section, both at DM thermal freeze out and in the current universe, and gives rise to a peculiar structure in peaks, that we recomputed in detail (Sec.~\ref{sec:sommerfeld}).

\medskip

In fig.~\ref{fig:summary} we compile the bounds in a summary chart: we shade away the intervals in mass which are excluded by each one of the gamma ray probes that we have considered and we contour with a dashed line the regions explorable with near future improvements. 

{\em The} MDM 5plet, which has a mass of 9.4 TeV as determined by a careful computation of its relic abundance (Sec.~\ref{sec:relic}), is severely tested by the Galactic Center line constraints from \HESS\ (sec.~\ref{sec:lines}): if the DM galactic profile is peaked like NFW, the model is ruled out; if the profile is cored like Burkert (or Isothermal), the model is still allowed. In this latter case, future {\sc Cta} line searches may enlarge the explored window but still are not expected to test that precise value in mass. In addition, future {\sc Cta} observations of the inner Galaxy are expected to have the power to explore the whole mass range $100\ {\rm GeV} \lesssim M_{\rm DM} \lesssim 30 \ {\rm TeV}$~\cite{Doro:2012xx, Silverwood:2014yza,Lefranc:2015pza} (applicable however only to non-flat profiles).

More generally, our results show that multi-TeV WIMPs can be significantly probed with Indirect Detection via gamma rays. 
Also, independently on the specific case of the 5plet with its peculiar roller-coaster cross section, the constraints in fig.~\ref{fig:bounds_continuum} (and fig.~\ref{fig:dwarves}) actually apply to any DM with $M_{\rm DM} \gtrsim 500$ GeV that annihilates into $W^+W^-$, $ZZ$, $Z\gamma$ or $\gamma\gamma$, as we have discussed above.

\medskip

Other channels of Indirect Detection for multi-TeV WIMP DM are possible, but are expected to have a somewhat shorter reach than the one we have considered here. The case of antiprotons is particularly interesting and has been explored intensively recently. Using the results in~\cite{Belanger:2012ta,Cirelli:2013hv,Cirelli:2014lwa,Boudaud:2014qra} and, in particular, the most recent assessment based on {\sc Ams-02} data~\cite{Giesen:2015ufa}, we can obtain constraints on the $W^+W^-$ cross section of Sec.~\ref{sec:sommerfeld}. We show the outcome in the summary chart as a shaded orange area: despite their relevance for smaller masses, these constraints do not change the global picture sketched by the gamma ray ones.

\medskip

In conclusion: Minimal Dark Matter, and more generally multi-TeV WIMP DM candidates, are arguably even more motivated than before, in the current context of absence of New Physics from the LHC. Gamma rays are a powerful probe for this class of models. {\em The} MDM 5plet candidate is ruled out or still allowed depending on the DM profile at the Galactic Center. Significant future progress is possible and may notably come from the observation of dwarf spheroidal galaxies.

\bigskip

\paragraph{Note:}
During the preparation of this work, we became aware of another group independently investigating gamma ray signals of Minimal Dark Matter scenarios~\cite{GILT}.

\bigskip

\small
\subsubsection*{Acknowledgments}
We thank Gabrijela Zaharijas for always very useful discussions! 
We also thank Brando Bellazzini, Alessandro Cuoco, Michael Gustafsson, Felix Kahlhoefer, Gary Mamon, Marco Nardecchia and Joe Silk.
M.C., F.S. and M.T. acknowledge the hospitality of the Institut d'Astrophysique de Paris ({\sc Iap}) while T.H. acknowledges that of the IPhT-Saclay.

\medskip

\footnotesize
\noindent Funding and research infrastructure acknowledgements: 
\begin{itemize}
\item[$\ast$] European Research Council ({\sc Erc}) under the EU Seventh Framework Programme (FP7/2007-2013)/{\sc Erc} Starting Grant (agreement n.\ 278234 --- `{\sc NewDark}' project),

\item[$\ast$] {\sc Erc} Advanced Grant 267117 (`{\sc Dark}') hosted by Universit\'e Pierre \& Marie Curie - Paris 6,

\item[$\ast$] French national research agency {\sc Anr} under contract {\sc Anr} 2010 {\sc Blanc} 041301,

\item[$\ast$] the {\sc Fnrs-Frs}, the {\sc Fria}, the {\sc Iisn}, an {\sc Ulb-Arc} and the Belgian Science Policy, IAP VI-11

\end{itemize}

\bigskip
\appendix

\footnotesize
\begin{multicols}{2}
  
\end{multicols}

\end{document}